\ifx\pdfoutput\undefined  
\documentclass[12pt, a4paper, twoside, dvips]{article}
\usepackage{color} \definecolor{webblue}{rgb}{0,0,0.5}
\else  
\documentclass[12pt, a4paper, twoside, pdftex]{article}
\usepackage{color} \definecolor{webblue}{rgb}{0,0,0.6}
\fi
\usepackage{theorem, bm}
\usepackage[colorlinks=true, linkcolor=webblue, citecolor=webblue,
 pagecolor=webblue, urlcolor=webblue, pdfstartview=FitBH,
 pdftitle={Gauge Transformations and Inverse Quantum Scattering with
 Medium-Range Magnetic Fields}, pdfauthor={Wolf Jung},
 bookmarksopen=true, bookmarksnumbered=true, pdfpagemode=None]{hyperref}
\theoremstyle{break}  \theorembodyfont{\itshape}
\newtheorem{thm}{Theorem}[section] \newtheorem{lem}[thm]{Lemma}
\newtheorem{dfn}[thm]{Definition} \newtheorem{prop}[thm]{Proposition}
\newtheorem{cor}[thm]{Corollary}
\theorembodyfont{\upshape}  \newtheorem{rmk}[thm]{Remark}
\newcommand{\be}{\begin{equation}} \newcommand{\ee}{\end{equation}}
\newcommand{\ba}{\begin{eqnarray*}} \newcommand{\ea}{\end{eqnarray*}}
\newcommand{\ban}{\begin{eqnarray}} \newcommand{\ean}{\end{eqnarray}}
\renewcommand{\hat}{\widehat} \renewcommand{\tilde}{\widetilde} 
 \renewcommand{\)}{\emph{)}}
\newcommand{\eref}[1]{\emph{\ref{#1}}}
\newcommand{\mybox}{\hspace*{\fill}\rule{2mm}{2mm}}
\DeclareSymbolFont{AMSb}{U}{msb}{m}{n}  
\DeclareSymbolFontAlphabet{\mathbb}{AMSb}
\def\C{\mathbb{C}}    \def\R{\mathbb{R}}
  \def\N{\mathbb{N}}
\def\div{\mathop{\rm div}\nolimits}  \def\arg{\mathop{\rm arg}\nolimits}
\def\tr{{\rm tr}}  \def\Ran{\mathop{\rm Ran}\nolimits}
\bmdefine\grad{\mathop{\rm grad}\nolimits}
\bmdefine\curl{\mathop{\rm curl}\nolimits}
\def\supp{\mathop{\rm supp}\nolimits} 
\newcommand{\e}[1]{\displaystyle {\rm e}^{\displaystyle #1}}
\renewcommand{\i}{\mathrm{i}}  
\newcommand{\eps}{\varepsilon}  
\newcommand{\dfrac}[2]%
 {\displaystyle\frac{\displaystyle #1}{\displaystyle #2}}
\renewcommand{\H}{\mathcal{H}}  \renewcommand{\O}{\mathcal{O}}
\renewcommand{\S}{\mathcal{S}}
\newcommand{\A}{\mathbf{A}}  \newcommand{\B}{\mathbf{B}}
\newcommand{\E}{\mathbf{E}}  \newcommand{\0}{\mathbf{0}}
\newcommand{\x}{\mathbf{x}}  \newcommand{\y}{\mathbf{y}}
\newcommand{\p}{\mathbf{p}}  \newcommand{\q}{\mathbf{q}}
\newcommand{\G}{\mathbf{G}}  \renewcommand{\u}{\mathbf{u}}
\renewcommand{\v}{\mathbf{v}}  \newcommand{\z}{\mathbf{z}}
\bmdefine{\bsigma}{\sigma}  \bmdefine{\bomega}{\omega}
\def\slim{\mathop{\rm s\!-\!lim}}  \def\wlim{\mathop{\rm w\!-\!lim}}
\newcommand{\loc}{{\mathrm{loc}}}
\date{Preprint of July 14, 2005.}
\author{Wolf Jung\\Stud.Ref.~at Studienseminar M\"onchengladbach\\
Abteistrasse 43--45, 41061 M\"onchengladbach, Germany\\
 \href{http://www.iram.rwth-aachen.de/~jung}%
  {http://www.iram.rwth-aachen.de/$\sim$jung}\\
 \href{mailto:jung@iram.rwth-aachen.de}{jung@iram.rwth-aachen.de}}
\title{Gauge Transformations and Inverse Quantum\\
Scattering with Medium-Range Magnetic Fields}
\begin{document} \maketitle
\begin{abstract}
\noindent  The time-dependent,  geometric method for high-energy
limits and inverse scattering is applied to nonrelativistic quantum
particles in external electromagnetic fields.  Both the Schr\"odinger-
and the Pauli equations in $\R^2$ and $\R^3$ are considered.  The
electrostatic potential $A_0$ shall be short-range,  and the magnetic
field $\B$ shall decay faster than $|\x|^{-3/2}$.  A natural class of
corresponding vector potentials $\A$ of medium range is introduced,
and the decay and regularity properties of various gauges are discussed,
including the transversal gauge, the Coulomb gauge,  and the Griesinger
vector potentials.  By a suitable combination of these gauges,
$\B$ need not be differentiable.  The scattering operator $S$ is not
invariant under the corresponding gauge transformations,  but
experiences an explicit transformation.  Both $\B$ and $A_0$ are
reconstructed from an X-ray transform,  which is obtained from the
high-energy limit of $S$.  Here previous results by Arians and Nicoleau
are generalized to the medium-range situation.  In a sequel paper,
medium-range vector potentials are applied to relativistic scattering.
\end{abstract}


\qquad\quad
2001 PACS:
02.30.Zz 

\newpage

\section{Introduction}
The scattering theory of a nonrelativistic quantum particle in an
electromagnetic field will be discussed under weak decay- and regularity
assumptions on the magnetic field. Consider first the corresponding
classical dynamics, i.e., the Lorentz force:
 \be \label{i1} m\ddot\x=e(\E+\dot\x\times\B) \ , \ee
where $m$ is the mass and $e$ is the charge of the particle, $\E(\x)$ is
electrostatic field strength, and $\B(\x)$ is the magnetostatic field.
More precisely,
$\B=\mu_0\mathbf{H}$ is the magnetic flux density, and $\mathbf{H}$ is the
magnetic field strength. The field is described in terms of a scalar
potential $A_0(\x)$ and a vector potential $\A(\x)$ according to
 $\E=-\grad A_0$ and $\B=\curl\A$. (Note that there is an alternative system
of units of measure in use, such that $\B=\mathbf{H}=c^{-1}\curl\A$, where
$c$ is the speed of light.) Now (\ref{i1}) is equivalent to a Hamiltonian
dynamic system with the Hamilton function
 \be \label{i2} H(\x,\,\p):=
 \frac1{2m}\,\Big(\p-e\A(\x)\Big)^2+eA_0(\x) \ , \ee
where $\p=m\dot\x+e\A(\x)$ is the canonical momentum. A nonrelativistic
quantum particle is described by a wave function
 $\psi(\x)\in L^2(\R^\nu,\,\C)$. Its time evolution is determined by the
Schr\"odinger equation $\i\hbar\dot\psi=H\psi$. The self-adjoint Hamiltonian
$H$ is given by
(\ref{i2}), with the canonical momentum operator $\p=-\i\hbar\nabla_\x\,$.
We set $\hbar=1$ and $e=1$. The Schr\"odinger operator is describing a
spin-$0$ particle, and the similar Pauli operator (\ref{ham1}) is describing
a particle of spin $1/2$, e.g., an electron. If $A_0(\x)$ and $\A(\x)$ decay
integrably as $|\x|\to\infty$, i.e., faster than $|\x|^{-1}$, then
$H$ is a short-range perturbation of $H_0:=\p^2/2m$. Its time
evolution is approximated by the free time evolution (generated by
$H_0$) as $t\to\pm\infty$, and the wave operators exist:
 \be \label{i3} \Omega_\pm\,\psi:=\lim_{t\to\pm\infty}
 \e{\i Ht}\,\e{-\i H_0t}\,\psi \ . \ee
Here the free state $\psi$ is an asymptotic state corresponding to the
scattering state $\Omega_\pm\,\psi$ as $t\to+\infty$ or $t\to-\infty$,
respectively. The scattering operator $S:=\Omega_+^*\Omega_-$ is mapping
incoming asymptotics to outgoing asymptotics. The vector potential $\A$ is
determined by the magnetic field $\B$ only up to a gradient. Under the gauge
transformation $\A'=\A+\grad\lambda$, the Hamiltonian $H$ is modified, but
the scattering operator $S$ is invariant if $\lambda(\x)\to0$ as
$|\x|\to\infty$.

If $\A$ does not decay integrably, the short-range wave operators
(\ref{i3}) need not exist. The time evolution generated by $H$ may be
described asymptotically in terms of long-range scattering theory, i.e., by
modifying the free time evolution. Loss and Thaller \cite{ltn, th} have
shown that the unmodified wave operators (\ref{i3}) still exist, if $\A(\x)$
is transversal, i.e., $\x\cdot\A(\x)=0$, and
$\A(\x)=\O(|\x|^{-(1/2+\delta)})$.
In (\ref{i2}), $(\A(\x))^2$ is short-range, but $\A(\x)\cdot\p$ is formally
long-range. It is effectively short-range, since $\A(\x)=-\x\times\G(\x)$
with $\G(\x)$ short-range, and $\A\cdot\p=\G\cdot\mathbf{L}$ with the
angular momentum $\mathbf{L}=\x\times\p$. This approach
generalizes to vector potentials $\A$ with the following property, which will
be called ``medium-range'': the transversal component
of $\A(\x)$, i.e., orthogonal to $\x$, is $\O(|\x|^{-(1/2+\delta)})$, and the
longitudinal component, i.e., parallel to $\x$, is decaying integrably.
--- The aims of the present paper are:
\begin{itemize}
 \item A class of medium-decay magnetic fields $\B$ is considered, such that
there is a medium-range vector potential $\A$ with $\curl\A=\B$ in $\S'$,
i.e., as a tempered distribution. The construction of $\A$ requires a decay
$\B(\x)=\O(|\x|^{-(3/2+\delta)})$ as in the case of the transversal gauge,
but the local regularity required of $\B$ can be weakened. The wave operators
are obtained analogously to \cite{ltn, elr}.
 \item The decay- and regularity properties of various gauges are discussed,
and the role of gauge transformations is emphasized: since the scattering
operator is not invariant in general under the substitution
$\A\to\A'=\A+\grad\lambda$ when $\A$ and $\A'$ are medium-range, we must
extract gauge-invariant quantities from $S$ (only these may be observed
in a physical experiment).
 \item The corresponding inverse scattering problem can be solved by
obtaining the X-ray transform of $\A$ from the high-energy limit of $S$.
This was done by Arians \cite{ar1} for short-range $\A$ under low
regularity assumptions, and by Nicoleau \cite{nfo} for smooth $\A$
of medium-range. Here these results are extended to low-regularity $\A$
of medium range. The inverse problem of relativistic
scattering with medium-range $\A$ is addressed in \cite{wjm2},
combining the techniques of \cite{ltr, th, wjde, wab}, including
obstacle scattering and the Aharanov--Bohm effect as well.
\end{itemize}
Only fields and particles in $\R^2$ and $\R^3$
are considered here. A measurable function $A_0:\R^\nu\to\R$ is a scalar
potential of \emph{short range}, if the multiplication operator $A_0(\x)$
is Kato-small with respect to $H_0=\frac1{2m}\p^2$, and if it
satisfies the Enss condition
 \be \label{vs} \int_0^\infty \Big\|
 \,A_0(\x)\,(H_0+\i)^{-1}\,F(|\x|\ge r) \,\Big\| \,dr \,<\,\infty \ , \ee
where $F(\dots)$ denotes
multiplication with the characteristic function of the indicated region.
This condition is satisfied, e.g., if $A_0\in L_\loc^2\,$, and if it decays
as $|\x|^{-\mu}$ with $\mu>1$. The magnetic field
 $\B:\R^\nu\to\R^{\nu'}$, $\nu':=\nu(\nu-1)/2$, corresponds to a $2$-form.

\begin{dfn}[Decay Conditions] \label{Dd}
$1$. Consider a magnetic field $\B:\R^3\to\R^3$ or $\B:\R^2\to\R$,
which is in $L^p(\R^\nu)$ for a $p>\nu$. It is of \emph{medium decay}, if it
satisfies a decay condition $|\B(\x)|\le C|\x|^{-\mu}$ for a $\mu>3/2$ and
large $|\x|$. In the case of $\R^3$, we also require that $\div\B=0$ in
$\S'$.

$2$. A vector potential $\A:\R^\nu\to\R^\nu$ is of \emph{medium range},
if it is continuous and satisfies $|\A(\x)|\le C|\x|^{-\mu}$ for some
$\mu>1/2$. In addition, the longitudinal part $\A(\x)\cdot\x/|\x|$ shall
decay integrably, i.e.,
 \be \int_0^\infty \sup\Big\{ \,|\A(\x)\cdot\x|/|\x|\,\, \Big|
 \,\,|\x|\ge r \,\Big\} \,dr \,<\,\infty \ . \ee
\end{dfn}

Note that $\A(\x-\x_0)$ is of medium range as well. The magnetic fields of
medium decay form a family of Banach spaces, cf.~Cor.~\ref{CB}.
By the decay at $\infty$, we have $\B\in L^p$ for a $p<2$ in addition.
The local regularity condition $p>\nu$ on $\B$ enables $\A$ to be continuous.
Since $\B$ need not be continuous, the case of an infinitely long solenoid is
included, where $\B:\R^2\to\R$ is the characteristic function of the
solenoid's cross section.
The \emph{Coulomb gauge} vector potential satisfies
$\div\A=0$, it is determined uniquely by $\curl\A=\B$ and $\A(\x)\to0$ as
$|\x|\to\infty$. (It is called transversal in QED, because
$\p\cdot\hat\A(\p)=0$.) The \emph{transversal gauge} vector potential
satisfies $\x\cdot\A(\x)=0$, it is determined uniquely by $\B$ if it is
continuous at $\x=0$. The \emph{Griesinger gauge} is introduced in
Sec.~\ref{subgr}, motivated by \cite{rg1}.

\begin{thm}[Vector Potentials] \label{Tg}
$1$. If $\A$, $\A'$ are medium-range vector potentials with
$\curl\A'=\curl\A$ in $\S'$, then there is a $C^1$-function
$\lambda:\R^\nu\to\R$ with $\A'=\A+\grad\lambda$. Moreover, the homogeneous
function $\Lambda(\x):=\lim_{r\to\infty} \lambda(r\x)$ exists and is
continuous for $\x\neq\0$.

$2$. If $\B$ is a magnetic field of medium decay, consider the vector
potential $\A$ with $\curl\A=\B$ in $\S'$, given in the Griesinger gauge.
It is of medium range, moreover it decays as $\O(|\x|^{-(\mu-1)})$,
if $\B$ decays as $\O(|\x|^{-\mu})$ and $3/2<\mu<2$. If $\B$ is continuous,
then the transversal gauge vector potential has the same properties.

$3$. Suppose that $\B$ satisfies a stronger decay condition with
$\mu>2$. Then the Coulomb gauge vector potential is of medium range, too.
Moreover, $\A$ is bounded by $C|\x|^{-1}$ in all of these gauges. In
$\R^3$, the flux of $\B$ through almost
every plane vanishes, and the Coulomb vector
potential is short-range. In $\R^2$, the flux $\Phi$ of $\B$ is finite, and
the Coulomb vector potential is short-range, iff $\Phi=0$.

$4$. If $\B$ is a magnetic field of medium decay, there is a special
choice of a medium-range vector potential $\A=\A^s+\A^r$ with $\curl\A=\B$
in $\S'$, where $\A^s$ is short-range and continuous, and $\A^r$ is
transversal and $C^\infty$, with
$|\,\partial_i\A_k^r(\x)|\le C|\x|^{-\mu}$, $\mu>1$. In addition,
$\div\A$ is continuous and decays integrably.
\end{thm}

Moreover, the Griesinger gauge $\A$ and the Coulomb gauge $\A$ are
regularizing, i.e., all $\partial_iA_k$ have the same local regularity as
$\B$. For the transversal gauge $\A$, $\partial_iA_k$ exists only as a
distribution in general if $\B$ is continuous. In \cite{th}, it is remarked
that the transversal gauge vector potential is better adapted to scattering
theory than the Coulomb gauge vector potential, if
$\B(\x)=\O(|\x|^{-(3/2+\delta)})$ and $\B$ is sufficiently regular.
The Coulomb gauge is superior in other cases:

\begin{rmk}[Advantages of Different Gauges] \label{Rac}
Suppose $\B:\R^\nu\to\R^{\nu'}$ is a magnetic field of medium decay with
$\B(\x)=\O(|\x|^{-\mu})$:

1. If $3/2<\mu\le2$, the Coulomb gauge vector potential is not of medium
range in general, and the wave operators (\ref{i3}) need not exist. The
transversal gauge vector potential may be used if $\B$ is continuous.
The Griesinger gauge works in any case.

2. If $\mu>2$, $\nu=2$, and $\int_{\R^2}\B\,dx\neq0$, then the Coulomb gauge
vector potential is of medium range as well, i.e., its longitudinal component
is decaying integrably. It is preferable to the transversal gauge because of
its better local regularity properties, and because the Hamiltonian
(\ref{i2}) is simplified due to $\p\cdot\A=\A\cdot\p$.

3. If $\mu>2$ and $\nu=3$, or $\nu=2$ and $\int_{\R^2}\B\,dx=0$, the Coulomb
gauge vector potential is short-range, but the transversal gauge or the
Griesinger gauge is short-range only in exceptional cases.
\end{rmk}

Under the assumptions of item~3, it is natural to use only short-range $\A$,
and the scattering operator $S$ is gauge-invariant. In the medium-range case,
we have a family of scattering operators, which are related by the
transformation formula (\ref{gts}):

\begin{thm}%
 [Gauge Transformation, Asymptotics, Inverse Scattering] \label{The}
Suppose that $\B$ is a magnetic field of medium decay in $\R^2$ or $\R^3$,
$\A$ is any medium-range vector potential with $\curl\A=\B$ in $\S'$, and
$\A_0$ is a short-range electrostatic potential according to
\emph{(\ref{vs})}.

$1$. For the Schr\"odinger- or Pauli operator $H$, the wave operators
$\Omega_\pm$ exist. 
Consider a gauge transformation $\A'=\A+\grad\lambda$ and
$\Lambda(\x)=\lim_{r\to\infty} \lambda(r\x)$ according to Thm.~\eref{Tg},
and denote the operators corresponding to $\A'$ by
$H'$, $\Omega_\pm'$, $S'$. The wave operators and scattering operators
transform under a change of gauge as
 \be \label{gts} \Omega_\pm' \,=\,
 \e{\i\lambda(\x)} \,\Omega_\pm\, \e{-\i\Lambda(\pm\p)} \qquad
 S' \,=\, \e{\i\Lambda(\p)} \,S\, \e{-\i\Lambda(-\p)} \ . \ee

$2$. Consider translations in momentum space by $\u=u\bomega$,
$\bomega\in S^{\nu-1}$. The scattering operator $S$ for the corresponding
Schr\"odinger- or Pauli equations in $\R^2$ and $\R^3$ has the asymptotics
 \be\label{hes} \slim_{u\to\infty} \,\e{-\i\u\x} \,S\, \e{\i\u\x} \,=\,
 \exp\Big\{ \i\int_{-\infty}^\infty \bomega\cdot\A (\x+\bomega t)
 \,dt \Big\} \ . \ee
$\B$ is reconstructed uniquely from the relative phase of this high-energy
limit of $S$. \(The absolute phase is not gauge-invariant, thus not
observable.\) Under stronger decay assumptions, error bounds and the
reconstruction of $A_0$ are given in Sec.~\eref{subrec}.
\end{thm}

Item~1 is due to \cite{rab} for $\nu=2$. Item~2 will be proved with the
time-dependent geometric method of Enss and Weder \cite{ew2}. It is due to
Arians
\cite{ar1, ard} for $\A$ of short range, and in addition for $\B:\R^2\to\R$
of compact support. The inverse scattering problem was solved before in
\cite{er} for $\A$ of exponential decay, and in \cite{nfo} for
$C^\infty$-$\A$ of medium range in the transversal gauge.
Analogous results for the Aharanov--Bohm
effect are discussed by Nicoleau \cite{nab} and Weder \cite{wab}.
The asymptotics for $C^\infty$-vector potentials of
medium range or long range are obtained in \cite{nab, ryab, yhs} using the
Isozaki--Kitada modification (cf.~Sec.~\ref{subik}).

\begin{rmk}[Gauge Invariance] \label{Rg}
1. Gauge freedom has two sides to it: we may choose a convenient gauge to
simplify a proof, but if the result depends on the gauge, it may be
insignificant from a physical point of view. Cf.~Sec.~\ref{subgi}.
Therefore existence and
high-energy limits of the wave operators are proved in
two steps: first by employing the nice properties of the special gauge
$\A=\A^s+\A^r$ according to item~4 of Thm.~\ref{Tg}, and then the result is
transfered to an arbitrary gauge by the transformation
(\ref{gts}). Thus (\ref{hes}) is valid in any
medium-range gauge, and the gauge-invariant relative phase is observable
in principle. Moreover, this approach shows that only the decay
properties of $\A$ are essential here, while the local regularity properties
are for technical convenience.

2. If $\nu=2$, $\mu>2$, and the flux of $\B$ is not vanishing, it is
possible to replace the medium-range techniques with short-range techniques
plus an adaptive gauge transformation, such that the vector potential is
decaying integrably in the direction of interest. Cf.~Cor.~\ref{Cc} and
\cite{ard, wjdip, wab}. In Sec.~\ref{subrec}, a different kind of adaptive
gauge transformation is used, such that the high-energy asymptotics of $H$
are simplified.
\end{rmk}

This paper is organized as follows: Vector potentials are discussed in
Sec.~\ref{Sg}, including the proof of Thm.~\ref{Tg}. In Sec.~\ref{Sx},
$\B$ is reconstructed from the X-ray transform of $\A$. The direct problem
of nonrelativistic scattering theory is addressed in Sec.~\ref{Sgo}.
Existence of the wave operators is proved in detail, because the same
techniques are needed later for the high-energy limit, but the reader is
referred to \cite{ltn, elr, ardip} for asymptotic completeness. Sec.~\ref{Si}
is dedicated to the inverse problem, and concluding remarks on
gauge invariance and on inverse scattering are given in Sec.~\ref{Sphint}.

\subsection*{Acknowledgment}
I wish to thank Silke Arians, Josef Bemelmans, Volker Enss,
Fernando Lled\'o, Olaf Post,
Christian Simader, Bernd Thaller, Ricardo Weder, and Dimitrij Yafaev
for inspiring discussions and useful hints.

\section{Fields and Gauges} \label{Sg}
To construct medium-range vector potentials, we are employing vector
analysis on $\R^2$ and $\R^3$ under low regularity
assumptions, controlling the decay at infinity. Some references to similar
results for Sobolev spaces over domains in $\R^\nu$ are included, and
\cite{dr} is a standard reference for vector analysis on manifolds
using distributional derivatives. The following notation will be employed
in the case of $\R^2$. It is motivated by identifying vectors and scalars in
$\R^2$ with vectors in $\R^3$,
 $\v=(v_1\,,\,v_2)^\tr \leftrightarrow (v_1\,,\,v_2\,,\,0)^\tr$
and $w \leftrightarrow (0\,,\,0\,,\,w)^\tr$:
 \ba \x\times\v := x_1v_2 - x_2v_1 &\qquad&
  \curl\v := \partial_1v_2 - \partial_2v_1 \\
  \x\times w := (x_2w\,,\,-x_1w)^\tr &\qquad&
  \curl w := (\partial_2w\,,\,-\partial_1w)^\tr \ . \ea

\subsection[Gauge Transformation of A]%
{Gauge Transformation of $\A$} \label{subga}
Suppose $\A:\R^\nu\to\R^\nu$ is a medium-range vector potential according
to Def.~\ref{Dd}. Assume $\curl\A=\0$ in $\S'$, i.e.,
 $\int\A\cdot\curl\phi\,dx=0$ for $\phi\in\S(\R^2,\,\R)$ or
$\phi\in\S(\R^3,\,\R^3)$, respectively. To show that $\A$ is a gradient,
one can mollify $\A$ and employ a density argument, or apply mollifiers
to the line integral over a closed curve. Here we shall use test
functions, and define $\lambda:\R^\nu\to\R$ by the Poincar\'e formula
for closed 1-forms,
 \be \label{linta} \lambda(\x) :=
 \int_0^1 \x\cdot\A(s\x)\,ds \ . \ee
For $\phi\in\S(\R^\nu,\,\R)$ and a unit vector $\bomega\in S^{\nu-1}$
consider
 \ban &&-\int_{\R^\nu} \lambda(\x)\,\bomega\cdot\nabla\phi(\x)\,dx \\[1mm]
 &=& -\int_{\R^\nu}\int_0^1
  \x\cdot\A(s\x)\,\bomega\cdot\nabla\phi(\x)\,ds\,dx \\[1mm]
 &=& \int_{\R^\nu}\int_0^1  \label{intacurl}
  \A(s\x)\cdot\curl\Big(\bomega\times\x\,\phi(\x)\Big)\,ds\,dx \;- \\[1mm]
 && -\int_{\R^\nu}\int_0^1 \bomega\cdot\A(s\x)\,  \label{secintla}
  \Big((\nu-1)\phi(\x)+\x\cdot\nabla\phi(\x)\Big)\,ds\,dx \ . \ean
This identity is verified with
 $\mathbf{a}\times(\mathbf{b}\times\mathbf{c})
 =(\mathbf{a}\cdot\mathbf{c})\mathbf{b}
 -(\mathbf{a}\cdot\mathbf{b})\mathbf{c}$.
Now (\ref{intacurl}) is vanishing by $\curl\A=\0$ in $\S'$. The substitution
$(s,\,\x)\to(s,\,\y)$ with $\y=s\x$ in (\ref{secintla}) gives
 \ban && -\int_{\R^\nu}\int_0^1 \bomega\cdot\A(\y)\,
  \Big((\nu-1)s^{-\nu}\phi(\y/s)
  +s^{-\nu-1}\y\cdot\nabla_\x\phi(\y/s)\Big)\,ds\,dy \\[1mm]
 &=& \int_{\R^\nu} \bomega\cdot\A(\y)\,
  \Big[s^{1-\nu}\phi(\y/s)\Big]_{s=0+}^1\,dy
 \;=\; \int_{\R^\nu} \bomega\cdot\A(\y)\,\phi(\y)\,dy \ . \ean
This shows $\A=\grad\lambda$ in $\S'$, and we have $\lambda\in C^1$ since
$\A$ is continuous, thus $\A=\grad\lambda$ pointwise.
Moreover, line integrals of $\A$ are path-independent. Now consider
 \be \Lambda(\x) := \lim_{r\to\infty} \lambda(r\x) =
 \lim_{r\to\infty} \int_0^r \x\cdot\A(t\x)\,dt \ . \ee
Since $\frac{\x}{|\x|}\cdot\A(\x)$ is short-range, convergence is uniform for
$|\x|\ge R$, thus $\Lambda$ is continuous on $\R^\nu\setminus\{\0\}$.
($\Lambda$ is $0$-homogeneous, and discontinuous at $\x=\0$ unless it is
constant.) If $\A$ is short-range, it is well-known that $\lambda$ can be
redefined such that $\lim_{|\x|\to\infty}\lambda(\x)=0$. This follows from
the estimate $\A(\x)=o(|\x|^{-1})$, see, e.g., \cite[Lemma~2.12]{wjdip}.
Item~1 of Thm.~\ref{Tg} is proved. \mybox

\begin{rmk}[Poincar\'e Lemma I] \label{RAL}
1. The same proof shows the following version of the Poincar\'e Lemma:
Suppose $\Omega\subset\R^\nu$, $p>\nu\ge2$, and
 $\A\in L_\loc^p(\Omega,\,\R^\nu)$ with
 $\int_\Omega A_i\partial_k\phi-A_k\partial_i\phi\,dx=0$ for all
 $\phi\in C_0^\infty(\Omega)$. If $\Omega$ is starlike around $\x=\0$,
define $\lambda:\Omega\to\R$ by (\ref{linta}) a.e., then $\lambda$ is
weakly differentiable with $\grad\lambda=\A$ almost everywhere.

2. If $\Omega$ is simply connected, then $\lambda$ is obtained piecewise.
For general $\Omega$, $\lambda$ need not exist globally. Now $\A$ is a
gradient, iff $\int_\Omega\A\cdot\v\,dx=0$ for all
 $\v\in C_0^\infty(\Omega,\,\R^\nu)$ with $\div\v=0$. Under this assumption
on $\A\in L_\loc^1(\Omega,\,\R^\nu)$, $\lambda$ is constructed by employing
mollifiers \cite{siso, gb}. For suitable $\Omega$, this result implies the
Helmholtz-Weyl decomposition of $L^p(\Omega,\,\R^\nu)$, $p>1$, which is
important, e.g., in fluid mechanics.
\end{rmk}

\subsection{The Transversal Gauge} \label{subtg}
Suppose that $\B:\R^3\to\R^3$ or $\B:\R^2\to\R$ is a magnetic field of medium
decay according to Def.~\ref{Dd}. The transversal gauge vector potential
$\A:\R^\nu\to\R^\nu$ is defined a.e.~by the Poincar\'e formula
for closed 2-forms,
 \be \label{tg} \A(\x) := -\x\times\int_0^1 s\B(s\x) \, ds \ . \ee

\begin{prop}[Transversal Gauge] \label{Ptg}
Suppose that $\B:\R^\nu\to\R^{\nu'}$ is a magnetic field of medium
decay with $\B(\x)=\O(|\x|^{-\mu})$ as
$|\x|\to\infty$, $\mu>3/2$.
In the transversal gauge, the vector potential $\A$ is defined by
\emph{(\ref{tg})}. Assume in addition that $\B$ is continuous. Then

$1.$ $\A$ is continuous and satisfies $\A(\x)=\O(|\x|^{-1})$ if $\mu>2$,
 $\A(\x)=\O(|\x|^{-1}\log|\x|)$ if $\mu=2$, and
 $\A(\x)=\O(|\x|^{-(\mu-1)})$ if $\mu<2$. Since $\x\cdot\A(\x)=0$ and
$\mu>3/2$, $\A$ is of medium range.

$2.$ We have $\curl\A=\B$ in $\S'$, but the weak partial derivatives
$\partial_iA_k$ and $\div\A$ do not exist in general.
\end{prop}

$\B$ is required to be continuous, to ensure that $\A$ is continuous.
(More generally,
$\B$ may have local singularities $c|\x-\x_0|^{-(1-\delta)}$, or a
jump discontinuity on a strictly convex line or surface, but a jump
discontinuity on a line through the origin is not permitted.)
The decay properties are given in \cite{ltn, ltr, th, elr} for $3/2<\mu<2$,
and in \cite{nab} for $\mu\neq2$. To achieve that $\A\in C^1$ with all
derivatives decaying integrably, we would have to assume $\B\in C^1$
with derivatives decaying faster than $|\x|^{-2}$.

\textbf{Proof}: 1. Define $b(r):=\sup_{|\x|=r} |\B(\x)|$ for $r\ge0$, then
$b(r)=\O(r^{-\mu})$ as $r\to\infty$. Now
 $|\A(\x)|\le |\x|^{-1} \int_0^{|\x|} r\,b(r)\,dr$ gives the desired
estimates (which are sharp).

2. Consider $\nu=3$ and a test function $\phi\in\S(\R^3,\,\R^3)$. We have
 \ban && +\int_{\R^\nu} \A(\x)\cdot\curl\phi(\x)\,dx \\[1mm]
 &=& -\int_{\R^\nu}\int_0^1  \label{intbatg}
  \Big(\x\times s\B(s\x)\Big)\cdot\curl\phi(\x)\,ds\,dx \\[1mm]
 &=& \int_{\R^3}\int_0^1  \label{intbcurl}
  s\B(s\x)\cdot\nabla\Big(\x\cdot\phi(\x)\Big)\,ds\,dx \\[1mm]
 && -\int_{\R^3}\int_0^1  \label{secintbatg}
  s\B(s\x)\cdot\Big(\phi(\x)+(\x\cdot\nabla)\phi(\x)\Big)\,ds\,dx \ , \ean
since
 $(\mathbf{a}\times\mathbf{b}) \cdot (\mathbf{c}\times\mathbf{d})
 =(\mathbf{a}\cdot\mathbf{c}) (\mathbf{b}\cdot\mathbf{d})
 -(\mathbf{a}\cdot\mathbf{d}) (\mathbf{b}\cdot\mathbf{c})$.
The integral (\ref{intbcurl}) is vanishing because of $\div\B=\0$ in $\S'$.
In (\ref{secintbatg}) we substitute $\y=s\x$ and obtain
 \ban && -\int_{\R^3}\int_0^1 \B(\y)\cdot \Big(s^{-2}\phi(\y/s)
  +s^{-3}(\y\cdot\nabla_\x)\phi(\y/s)\Big)\,ds\,dy \\[1mm]
 &=& \int_{\R^3} \B(\y)\cdot
  \Big[s^{-1}\phi(\y/s)\Big]_{s=0+}^1\,dy
 \;=\; \int_{\R^\nu} \B(\y)\cdot\phi(\y)\,dy \ . \ean
In dimension $\nu=2$, we have $\phi\in\S(\R^2,\,\R)$, and (\ref{intbatg})
equals
 \ban && -\int_{\R^2}\int_0^1 s\B(s\x)\,\x\cdot\nabla\phi(\x)\,ds\,dx \\[1mm]
 &=& -\int_{\R^2}\int_0^1 \B(\y)\,s^{-2}\y\cdot\nabla_\x\phi(\y/s)\,ds\,dy
 \label{intgtnu2}
 \;=\; \int_{\R^2} \B(\y)\,\Big[\phi(\y/s)\Big]_{s=0+}^1\,dy \ . \ean
Thus $\curl\A=\B$ in $\S'(\R^3,\,\R^3)$ or $\S'(\R^2,\,\R)$, respectively.
--- For $\nu=2$, suppose that
 $\B(r\cos\theta,\,r\sin\theta)=(1+r)^{-\mu} f(\theta)$, where $f$ is
singular continuous. Then none of the weak derivatives $\partial_iA_k$ or
$\div\A$ exists in $L_\loc^1(\R^2)$.
In the following example, the derivatives exist but they are not short-range:
 $\B(r\cos\theta,\,r\sin\theta)=(1+r)^{-\mu} \cos(r^\mu \theta)$.
Similar examples are constructed in $\R^3$. (The condition $\div\B=0$ is
satisfied, e.g., by $\B(\x)=\x\times\grad g(\x)$.) \mybox

\begin{rmk}[Poincar\'e Lemma II] \label{RBA}
1. Suppose $\Omega\subset\R^3$, $p>3/2$,
 and $\B\in L_\loc^p(\Omega,\,\R^3)$ with
 $\int_\Omega\B\cdot\grad\phi\,dx=0$ for all $\phi\in C_0^\infty(\Omega)$.
If $\Omega$ is starlike around $\x=\0$, define $\A(\x)$ by (\ref{tg}) a.e.,
then the same proof shows
 $\int_\Omega\A\cdot\curl\phi\,dx=\int_\Omega\B\cdot\phi\,dx$ for
 $\phi\in C_0^\infty(\Omega,\,\R^3)$. This vector potential is not
weakly differentiable in general.

2. On an arbitrary domain $\Omega$, a vector potential $\A$ exists if
 $\int_\Omega\B\cdot\phi\,dx=0$ for all $\phi\in C_0^\infty(\Omega,\,\R^3)$
with $\curl\phi=\0$. If $C_0^\infty$ can be replaced with $C^\infty$,
then $\A$ can be chosen such that it vanishes at the (regular) boundary
$\partial\Omega$: $\A$ is constructed in \cite{ww} by potential theory,
and in \cite{rg1} by a mollified version of (\ref{tg}), see below. These
vector potentials are weakly differentiable with
 $\|\partial_iA_k\|_p\le c_p\|\B\|_p\,$, $1<p<\infty$.
\end{rmk}

\subsection{The Griesinger Gauge} \label{subgr}
For a magnetic field $\B:\R^\nu\to\R^{\nu'}$ of medium decay, the Griesinger
gauge vector potential shall be defined by employing a mollifier
$h\in C_0^\infty(\R^\nu,\,\R)$ with $\int_{\R^\nu}h\,dx=1$:
 \ban \A(\x) &:=& \label{gr1}
 -\int_{\R^\nu}\int_0^1 h(\z)\,(\x-\z)\times s\B(s\x+(1-s)\z)\,ds\,dz \\[1mm]
 &=& -\int_{\R^\nu}\int_1^\infty  \label{gr2}
 h(\x-t(\x-\y))\,t^{\nu-2}(t-1)\,(\x-\y)\times\B(\y)\,dt\,dy \ . \ean
(\ref{gr1}) looks like a mollified version of the transversal gauge
(\ref{tg}), which is recovered formally for $h(\z)\to\delta(\z)$. Note that
$\B$ is averaged over a ball of radius $\simeq(1-s)$, which is shrinking to
a point as $s\to1$ in (\ref{gr1}), or $\y\to\x$ in (\ref{gr2}). It turns out
that the integral kernel in the latter equation is weakly singular.

This definition is the exterior domain analog to the construction found
by Griesinger for interior domains \cite{rg1}, i.e.,
 $\int_1^\infty ds$ was replaced with $-\int_0^1ds$. (Her original
construction
is less suitable for magnetic fields of medium decay, because it would
require $\B(\x)=\O(|\x|^{-2-\delta})$, and vanishing flux in the case of
$\R^2$.) The technique goes back to Bogovskij's solution of $\div\v=f$
\cite{bogo, bs, gb}. In the case of a bounded domain, the vector field
satisfies
 $\|\partial_iv_k\|_p\le c_p\|f\|_p\,$ or
 $\|\partial_iA_k\|_p\le c_p\|\B\|_p\,$, respectively ($1<p<\infty$).
--- Item~2 of Thm.~\ref{Tg} is contained in the following

\begin{prop}[Griesinger Gauge] \label{Ptgr}
Fix $h\in C_0^\infty(\R^\nu,\,\R)$ with $\int_{\R^\nu}h\,dx=1$. Suppose
$\B$ is a magnetic field of medium decay with $\B(\x)=\O(|\x|^{-\mu})$,
$\mu>3/2$. The Griesinger gauge vector potential $\A$ is defined by
\emph{(\ref{gr1})}.

$1.$ $\A$ is continuous and satisfies $\A(\x)=\O(|\x|^{-1})$ if $\mu>2$,
$\A(\x)=\O(|\x|^{-1}\log|\x|)$ if $\mu=2$, and
$\A(\x)=\O(|\x|^{-(\mu-1)})$ if $\mu<2$. The longitudinal component of
$\A$ is short-range, thus $\A$ is of medium range.

$2.$ $\A$ has weak partial derivatives in $L_\loc^2(\R^\nu)$, and
$\curl\A=\B$ almost everywhere. But the weak derivatives do not decay as
$\O(|\x|^{-\mu})$ in general.
\end{prop}

\textbf{Proof}: 1. Choose $p>\nu$ with $\B\in L^p(\R^\nu,\,\R^{\nu'})$,
$q:=1/(1-1/p)$, and fix $R>0$ such that $h(\z)=0$ for $|\z|\ge R$ and
 $|\B(\x)|\le c(1+|\x|)^{-\mu}$ for $|\x|\ge R$. By H\"older's inequality
we have
 \be\label{gr3} |\A(\x)| \le (|\x|+R)\, \|h\|_q
 \int_0^1 s\|\B(s\x+(1-s)\z)\,\chi(|\z|\le R)\|_p\,ds \ . \ee
Note that the norm of $\B$ is considered on the ball of radius $(1-s)R$
around $s\x$. For $|\x|\le2R$ consider
 $\|\dots\|_p\le\|\z\mapsto\B(s\x+(1-s)\z)\|_p=(1-s)^{-\nu/p}\|\B\|_p\,$,
thus
 \be |\A(\x)| \le 3R\, \|h\|_q \,\|\B\|_p \int_0^1 s(1-s)^{-\nu/p}\,ds \ee
is bounded. For $|\x|\ge2R$, the $s$-interval in (\ref{gr3}) is split:
{\renewcommand{\labelenumi}{\alph{enumi})}\begin{enumerate}
\item For $\displaystyle 0\le s\le\frac{2R}{|\x|+R}$ we have $1-s\ge1/3$,
 thus $\|\dots\|_p\le3^{\nu/p}\|\B\|_p=:c_1\,$.
\item For $\displaystyle \frac{2R}{|\x|+R}\le s\le1$ we have
 $|s\x+(1-s)\z|\ge R$ and $|s\x+(1-s)\z|\ge s|\x|/2$, therefore
 $\|\dots\|_p\le c_2(1+s|\x|)^{-\mu}$.
\end{enumerate}}
Now $\|\B(s\x+(1-s)\z)\,\chi(|\z|\le R)\|_p\le c_3(1+s|\x|)^{-\mu}$ for
 $0\le s\le1$, and (\ref{gr3}) yields the desired estimate for $|\A(\x)|$,
which is optimal. The stronger bound for the longitudinal component is
obtained by replacing the factor $(|\x|+R)\to R$ in (\ref{gr3}).

Write (\ref{gr2}) as $\A(\x)=\int_{\R^\nu} G(\x,\,\x-\y)\,\B(\y)\,dy$.
The kernel $G(\x,\,\z)$
is bounded by $c\,(1+|\x|^{\nu-1})\,|\z|^{-(\nu-1)}$, analogously to
(\ref{gr7}) below. Thus it is weakly singular, and $\B\in L^p$ implies
that $\A$ is continuous (adapting Thm.~II.9.2 in \cite{gb}).

2. We show $\curl\A=\B$ in $\S'$ by the techniques from Sec.~\ref{subtg},
and establish the higher regularity of the Griesinger gauge afterwards
(which was not possible for the transversal gauge). Consider $\nu=3$ and
$\phi\in\S(\R^3,\,\R^3)$. As in (\ref{intbcurl}) and (\ref{secintbatg})
we have
 \ban && +\int_{\R^3} \A(\x)\cdot\curl\phi(\x)\,dx \\[1mm]
 &=& \int_{\R^3}\int_{\R^3}\int_0^1  \label{gr4}
  h(\z)\,s\B(s\x+(1-s)\z)\cdot\nabla_\x
  \Big( (\x-\z)\cdot\phi(\x)\Big)\,ds\,dz\,dx \\[1mm]
 &-& \int_{\R^3}\int_{\R^3}\int_0^1  \label{gr5}
  h(\z)\,s\B(s\x+(1-s)\z)\cdot
  \Big(\phi(\x)+((\x-\z)\cdot\nabla_\x)\phi(\x)\Big)\,ds\,dz\,dx \ .
  \rule{6mm}{0mm} \ean
In (\ref{gr4}), the $\x$-integral is vanishing for a.e.~$s$ and $\z$,
because $\div\B=\0$ in $\S'$. In (\ref{gr5}), the substitution
$(s,\,\z,\,\x)\to(s,\,\z,\,\y)$ with $\y=s\x+(1-s)\z$ yields
 \ban && -\int_{\R^3}\int_{\R^3}\int_0^1 h(\z)\,\B(\y)\cdot \nonumber\\[1mm]
 && \cdot\Big(s^{-2}\phi(\z+(\y-\z)/s)
  +s^{-3}((\y-\z)\cdot\nabla_\x)\phi(\z+(\y-\z)/s)\Big)\,ds\,dz\,dy \\[1mm]
 &=& \int_{\R^3}\int_{\R^3} h(\z)\,\B(\y)\cdot
  \Big[s^{-1}\phi(\z+(\y-\z)/s)\Big]_{s=0+}^1\,dz\,dy \\[1mm]
 &=& \int_{\R^3}h(\z)\,dz\int_{\R^3} \B(\y)\cdot\phi(\y)\,dy \ , \ean
thus $\curl\A=\B$ in $\S'(\R^3,\,\R^3)$. For $\nu=2$ we obtain
$[\phi(\z+(\y-\z)/s)\Big]_{s=0+}^1=\phi(\y)$ analogously to
(\ref{intgtnu2}). (The same technique works for
$\int_1^\infty ds$, but for $\nu=2$ we have
 $[\phi(\z+(\y-\z)/s)\Big]_{s=\infty-}^1=\phi(\y)-\phi(\z)$, thus
$\curl\A(\x)=\B(\x)-h(\x)\int_{\R^2}\B(\y)\,dy$.)

The existence of all weak derivatives $\partial_iA_k$,
 $1\le i,\,k\le\nu$, is shown first for $\tilde\B\in C_0^\infty$ (without
restriction on $\div\tilde\B$), and then a density argument covers the
general case. To simplify the notation, we consider only $\div\A$:
(\ref{gr2}) implies, as a principal value,
 \ban \div\tilde\A(\x) \label{gr6}
 &=&\int_{\R^\nu} K(\x,\,\x-\y)\cdot\tilde\B(\y)\,dy \qquad\mbox{with}\\[1mm]
 K(\x,\,\z) &:=&
 -\z\times\int_1^\infty\nabla h(\x-t\z)\,t^{\nu-2}(t-1)^2\,dt \\[1mm]
 &=& -\frac\z{|\z|^{\nu+1}}\times\int_{|\z|}^\infty
 \nabla h\Big(\x-r\frac\z{|\z|}\Big)\,r^{\nu-2}(r-|\z|)^2\,dr \ . \ean
The most singular contribution is
 \be \label{gr7}
 -\frac\z{|\z|^{\nu+1}}\times\int_0^\infty
 \nabla h\Big(\x-r\frac\z{|\z|}\Big)\,r^\nu\,dr
  = \O\Big((1+|\x|^\nu)|\z|^{-\nu}\Big)\ . \ee
The Calder\'on--Zygmund Theorem \cite{cz}, cf.~\cite{rg1, gb}, shows that the
principal value integral (\ref{gr6}) is well-defined, and
 \be\label{gr8} \|(1+|\x|)^{-\nu}\div\tilde\A\|_p\le c_p\|\tilde\B\|_p \ .\ee
Approximating the given $\B\in L^p$ with $\tilde\B\in C_0^\infty$,
(\ref{gr8}) shows $\div\A\in L_\loc^p\,$.

Now $\partial_iA_k\in L_\loc^p(\R^\nu)$ is shown analogously,
and $\curl\A=\B$ in $\S'$ implies
$\curl\A=\B$ a.e.. I do not know if~$\partial_iA_k\in L^p(\R^\nu)$, or if
these weak derivatives decay integrably as $|\x|\to\infty$ (short-range
terms). But assuming that all $\partial_iA_k(\x)=\O(|\x|^{-\mu})$,
i.e., having the
same decay as $\B(\x)$, would imply $\A(\x)=\O(|\x|^{-(\mu-1)})$, which is
not true in general if $\mu>2$. \mybox

\begin{cor}[Banach Spaces] \label{CB}
For $R>0$, $p>\nu$, and $\mu>3/2$, define a norm
 \be \|\B\|_{R,p,\mu} := \|\B(\x)\,\chi(|\x|\le R)\|_p
 + \|\,|\x|^\mu\,\B(\x)\,\chi(|\x|\ge R)\|_\infty \ , \ee
and denote by $\mathcal{M}_{R,p,\mu}$ the Banach space of magnetic
fields with finite norm, and with $\div\B=0$ in $\S'$ if $\nu=3$.
The vector space of magnetic fields with medium decay is the union of
these Banach spaces. For fixed $h\in C_0^\infty$,
the proof of item~$1$ above shows that
the Griesinger gauge is a bounded operator
 $\mathcal{M}_{R,p,\mu}\to C^0(\R^\nu,\,\R^\nu)$.
\end{cor}

\subsection{The Coulomb Gauge}
In the Coulomb gauge, the vector potential $\A$ is defined by
 \be \label{coco} \A(\x) := -\frac1{\omega_\nu}\,
 \int_{\R^\nu}\dfrac{\x-\y}{|\x-\y|^{\,\nu}} \,\times\,\B(\y)\,dy \ee
with $\omega_2:=|S^1|=2\pi$ and $\omega_3:=|S^2|=4\pi$.

\begin{prop}[Coulomb Gauge] \label{PC}
Suppose $\B:\R^\nu\to\R^{\nu'}$ is a magnetic field of medium decay with
$\B(\x)=\O(|\x|^{-\mu})$, $\mu>3/2$. The Coulomb gauge vector potential
$\A:\R^\nu\to\R^\nu$ is defined by \emph{(\ref{coco})}.

$1.$ $\A$ is continuous and weakly differentiable with
 $\|\partial_iA_k\|_2\le\|\B\|_2\,$. It satisfies $\div\A=0$ and
$\curl\A=\B$ a.e..

$2.$ If $\mu>2$, then $\A$ is of medium range. For $\nu=3$ it is short-range.
For $\nu=2$ it is short-range, iff the flux is vanishing:
$\int_{\R^2}\B\,dx=0$.

$3.$ If $3/2<\mu\le2$, then $\A$ satisfies $\A(\x)=\O(|\x|^{-(\mu-1)})$, or
$\A(\x)=\O(|\x|^{-1}\log|\x|)$ if $\mu=\nu=2$. But the longitudinal component
of $\A$ does not decay integrably in general, and $\A$ need not be of
medium range.
\end{prop}

Thus the Coulomb  $\A$ has better differentiability properties than the
transversal gauge, even if $\B$ is continuous. The decay properties are
better for $\mu>2$ (and vanishing flux), but not sufficient in general
if $\mu\le2$. --- In \cite{nww}, $|\A(\x)|$ is estimated in $\R^3$ and in
exterior domains. We shall employ a convolution of Riesz potentials:

\begin{lem}[Riesz Potentials] \label{Lrp}
For $\nu\in\N$ and $0<\alpha,\,\beta<\nu$ with $\alpha+\beta>\nu$,
we have the convolution on $\R^\nu$
 \be |\x|^{-\alpha} \,*\, |\x|^{-\beta} \,=\,
 C_{\alpha,\,\beta;\,\nu} \, |\x|^{-(\alpha+\beta-\nu)} \ . \ee
\end{lem}

This formula is obtained form an elementary scaling argument and convergence
proof. The constant is determined in \cite[p.~136]{he}.

\textbf{Proof} of Prop.~\ref{PC}: 1. $\A$ is continuous since the integral
kernel in (\ref{coco}) is weakly singular and $\B\in L^p$ for some $p>\nu$,
cf.~Thm.~II.9.2 in \cite{gb} (the condition of a bounded domain is overcome
by splitting $\B$). The convolution (\ref{coco}) is obtained by
differentiating the fundamental solution of the Laplacian: we have
$\A=\curl\mathbf{U}$ with $-\Delta\mathbf{U}=\B$. This implies $\div\A=0$
and $\curl\A=\B$ in $\S'$. The
Fourier transforms satisfy $\i\p\cdot\hat\A(\p)=0$ and
$\i\p\times\hat\A(\p)=\hat\B(\p)$, thus
$\p^2\hat\A(\p)=\i\p\times\hat\B(\p)$. It remains to show that $\A$ is
weakly differentiable. Now
$|\widehat{\partial_iA_k}(\p)|\le|\hat\B(\p)|$ a.e., thus
 $\|\partial_iA_k\|_2\le\|\B\|_2$. We have $\B\in L^p$ for $p_1<p\le p_2\,$,
with $p_1<2$ and $p_2>\nu$. By the Calder\'on--Zygmund Theorem
\cite{cz, rg1, gb},
 $\|\partial_iA_k\|_p\le c_p\|\B\|_p$ for $p_1<p\le p_2\,$.

2. Now $\mu>2$, and we may assume $2<\mu<3$. Split $\B=\B^{(1)}+\B^{(2)}\,$,
such that $\B^{(1)}\in L^p$ has compact support, with
 $\int_{\R^2}\B^{(1)}\,dx=0$ in the case of $\R^2$, and such that
 $|\B^{(2)}(\x)|\le c|\x|^{-\mu}$ for $\x\in\R^\nu$
($\div\B^{(i)}=0$ is not required). Split $\A=\A^{(1)}+\A^{(2)}$
according to (\ref{coco}), i.e., by applying the convolution to $\B^{(i)}$
individually. If $\nu=3$, we have
$|\A^{(1)}(\x)|=\O(|\x|^{-2})$, since $|\x-\y|^{-2}=\O(|\x|^{-2})$ for
$\y\in\supp(\B^{(1)})$, and $|\A^{(2)}(\x)|=\O(|\x|^{-(\mu-1)})$ by
Lemma~\ref{Lrp}, thus $\A$ is short-range. For $\nu=2$ we claim
 \be \label{nuts} \A(\x) = \dfrac1{2\pi|\x|^2}\,
 \left(\begin{array}{c} -x_2 \\[1mm] x_1 \end{array}\right)
 \int_{\R^2}\B(\y)\,dy \,+\, \O(|\x|^{-(\mu-1)}) \ . \ee
The integral kernel of (\ref{coco}) is decomposed as follows:
 \[ -\dfrac{\x-\y}{|\x-\y|^2} = -\dfrac{\x}{|\x|^2}
 + \dfrac{(\x\times\y)\times(\x-\y)-(\x\cdot\y)(\x-\y)}{|\x|^2|\x-\y|^2}
  = -\dfrac{\x}{|\x|^2}
 + \O\Big(\,\dfrac{|\y|}{|\x|\,|\x-\y|} \,\Big)  \]
When applying this kernel to $\B^{(1)}\,$, the first integral is vanishing and
the second is $\O(|\x|^{-2})$. Applying it to $\B^{(2)}\,$, the first integral
gives the leading term in (\ref{nuts}), and the second integral is bounded by
 \be c\int_{\R^2} \dfrac{|\y|}{|\x|\,|\x-\y|}\,|\y|^{-\mu}\,dy
 = \dfrac{c}{|\x|}\int_{\R^2} |\x-\y|^{-1}\,|\y|^{-(\mu-1)}\,dy
 = c'|\x|^{-(\mu-1)} \ee
by Lemma~\ref{Lrp}. Thus (\ref{nuts}) is proved. (For $\B$ of compact support
this is due to \cite{tss}.) If the flux of $\B$ is
vanishing, then $\A$ is short-range. If not, then $\A$ is still of medium
range, since the leading term is transversal.

3. Now $3/2<\mu\le2$. Split $\B$ and $\A$ as in the previous item, then
$\A^{(1)}(\x)=\O(|\x|^{-2})$, $|\B^{(2)}(\x)|\le c|\x|^{-\mu}$, and
Lemma~\ref{Lrp}
gives $|\A^{(2)}(\x)|\le c'|\x|^{-(\mu-1)}$, except for the case of
$\mu=\nu=2$: then compute the bound explicitly for $c(1+|\x|^2)^{-1}$.
Consider the vector potential $\A(\x)$ given by
 \[ \dfrac1{(|\x|^2+1)^2}\,\left(\begin{array}{c}
 x_1(x_1^2-x_2^2+1) \\[1mm] x_2(x_1^2-x_2^2-1) \end{array}\right)
 \quad\mbox{or}\quad
 \dfrac1{(|\x|^2+1)^2}\,\left(\begin{array}{c}
 x_1(x_1^2-x_2^2+x_3^2+1) \\[1mm]
 x_2(x_1^2-x_2^2-x_3^2-1) \\[1mm] 0 \end{array}\right) \ , \]
respectively, and define $\B:=\curl\A$. We have $\B\in C^\infty$ with
$\B(\x)=\O(|\x|^{-2})$ and $\div\A=0$, thus $\A$ is the Coulomb gauge vector
potential for the medium-decay $\B$. Now
$A_1(x_1,\,0)=\dfrac{x_1}{x_1^2+1}$ or
$A_1(x_1,\,0,\,0)=\dfrac{x_1}{x_1^2+1}$, respectively, shows that the
longitudinal component is not short-range, and $\A$ is not of
medium range. \mybox

In the last example, $\Lambda(\x):=\lim_{r\to\infty} \lambda(r\x)$
according to item~1 of Thm.~\ref{Tg} does not exist for the gauge
transformation from transversal gauge to Coulomb gauge. The scattering
operator exists for the transversal gauge, but not for the Coulomb gauge.
Note that for rotationally symmetric $\B:\R^2\to\R$, the Coulomb gauge agrees
with the transversal gauge. Therefore, this vector potential combines the
regularity properties of the Coulomb gauge with the decay properties of
the transversal gauge: it is weakly differentiable and
of medium range for all $\mu>3/2$.

If $\mu>2$, the flux of $\B$ is finite if $\nu=2$, and for $\nu=3$, the flux
through almost every plane is vanishing. When $\nu=2$ and
 $\int_{\R^2}\B\,dx\neq0$, consider the natural class of medium-range vector
potentials satisfying $\A(\x)=\O(|\x|^{-1})$. The corresponding gauge
transformations $\lambda(\x)$ have the property that $\Lambda(\x)$ is
Lipschitz continuous for $|\x|>\eps$. This class contains the Coulomb gauge,
and the transversal gauge as well if $\B$ is continuous.

\begin{cor}[Adaptive Gauges] \label{Cc}
Suppose $\nu=2$ and $\B$ is of medium decay with $\B(\x)=\O(|\x|^{-\mu})$
for a $\mu>2$, and $\Phi:=\int_{\R^2}\B\,dx\neq0$. For any direction
 $\bomega\in S^1$ there is a vector potential $\A^\bomega$ of medium range,
such that $\A^\bomega(\x)=\O(|\x|^{-1})$, and $\A^\bomega(\x)$ decays
integrably as $\O(|\x|^{-(\mu-1)})$ within sectors around $\pm\bomega$.
Moreover, $\div\A^\bomega$ is continuous with
 $\div\A^\bomega(\x)=\O(|\x|^{-2})$.
\end{cor}

\textbf{Proof}: Denote the Coulomb gauge vector potential by $\A$ and
observe (\ref{nuts}). Choose a $2\pi$-periodic function $f\in C^2(\R,\,\R)$,
such that $f'(\theta)\equiv\Phi/(2\pi)$ for $\theta$ in intervals around
$\arg(\pm\bomega)$. Choose $\lambda\in C^2(\R^2,\,\R)$ with
 $\lambda(\x)\equiv f(\arg\x)$ for large $|\x|$. Then consider
 $\A^\bomega:=\A-\grad\lambda$. \mybox

Similar constructions are found, e.g., in \cite{wab}. If $\B$ has
compact support, $\A^\bomega$ can be chosen to vanish in these sectors for
large $|\x|$: this is achieved by subtracting a gradient, or by a
shifted transversal gauge if $\B$ is continuous \cite{ard, wjdip, wab}.

\subsection[The H\"ormander Decomposition of A]%
{The H\"ormander Decomposition of $\A$} \label{subd}
The following lemma is a special case of \cite[Lemma~3.3]{hlm}, which
is a standard tool in long-range scattering theory. (It is used
to improve the decay of derivatives of the long-range part $A_0^l\,$,
where $A_0=A_0^s+A_0^l\,$.)

\begin{lem}[H\"ormander Decomposition] \label{Lh}
Suppose that $V\in C^1(\R^\nu)$ and $V(\x)=\O(|\x|^{-m_0})$,
$\partial^\gamma V(\x)=\O(|\x|^{-m_1})$ for $|\gamma|=1$,
with $m_0\ge m_1-1>0$. For $0<\Delta<\min(1,\,m_1-1)$ there is a
decomposition $V=V_1+V_2$ such that: $V_1\in C^1$ is a short-range potential
with $V_1(\x)=\O(|\x|^{-\lambda})$ for $\lambda=\max(m_0,\,m_1-\Delta)>1$,
and $V_2\in C^\infty$ satisfies $\partial^\gamma V_2(\x)=\O(|\x|^{-m_j'})$
for $|\gamma|=j$, $j\in\N_0\,$. Here $m_0'=m_0$ and
$m_j'=\max(m_0+j\Delta,\,m_1+(j-1)\Delta)$ for $j\in\N$.
\end{lem}

For a given medium-decay $\B$ we want to obtain a corresponding medium-range
$\A=\A^s+\A^r$, such that $\A^s$ is short-range, $\A^r$ is transversal with
short-range derivatives, and $\div\A$ is short-range.
The transversal gauge does not
satisfy our requirements because it need not be differentiable, and the
Coulomb gauge is weakly differentiable and satisfies the condition on
$\div\A$, but its longitudinal part need not decay integrably. For
the Griesinger gauge, I do not know how to control the decay of the
derivatives. Lemma~\ref{Lh} cannot be applied directly to $\B$ or to a
known $\A$, because it requires $\B\in C^1$ or $\A\in C^1$, respectively.

\textbf{Proof} of Thm.~\ref{Tg}, item~4:
Suppose $\B(\x)=\O(|\x|^{-\mu})$ for $|\x|\to\infty$ and consider the Coulomb
gauge vector potential $\A^c$ according to (\ref{coco}). If $\mu>2$, we may
take $\A^s:=\A^c$ and $\A^r:=\0$, except in the case of $\nu=2$ and
$\int\B\,dx\neq0$: then $\A^r$ equals the first term in (\ref{nuts}) for
large $|\x|$. Thus we may assume
$3/2<\mu<2$, or
$\mu=3/2+3\delta$ with $0<\delta<1/6$. $\A^c$ is continuous, weakly
differentiable, and satisfies $\A^c(\x)=\O(|\x|^{-1/2-3\delta})$.
Choose a function $\eta\in C^\infty(\R^\nu,\,\R)$ with $\eta(\x)=0$ for
$|\x|<1$ and $\eta(\x)=1$ for $|\x|>2$, and consider the decomposition
$\A^c=\A^{(0)}+\A^\infty\,$:
 \ba\A^{(0)}(\x) &=& -\frac1{\omega_\nu}\,\int_{\R^\nu}(1-\eta(\x-\y))\,
 \dfrac{(\x-\y)}{|\x-\y|^{\,\nu}}\,\times\,\B(\y)\,dy \ , \\[2mm]
 \A^\infty(\x) &=& -\frac1{\omega_\nu}\, \int_{\R^\nu}\eta(\x-\y)\,
 \dfrac{(\x-\y)}{|\x-\y|^{\,\nu}} \,\times\,\B(\y)\,dy \ . \ea
Now $\A^{(0)}$ is short-range and $\A^\infty\in C^\infty$ with
$\A^\infty(\x)=\O(|\x|^{-1/2-3\delta})$. The derivatives are given by
convolutions $\partial_i A^\infty_k=\sum K_{ikl}*B_l$, where
$K_{ikl}(\x)=\O(|\x|^{-\nu})$ as $|\x|\to\infty$ and $K_{ikl}(\x)=0$ for
$|\x|<1$. Thus $|K_{ikl}(\x)|\le c|\x|^{-(\nu-\delta)}$ for $\x\in\R^\nu$.
Split $\B=\B^{(1)}+\B^{(2)}$, such that $\B^{(1)}$ has compact support
and $|\B^{(2)}(\x)|\le c'|\x|^{-3/2-3\delta}$. The estimate
$\partial_i A^\infty_k(\x)=\O(|\x|^{-3/2-2\delta})$ is obtained from
Lemma~\ref{Lrp}. A medium-range $\A$ can be constructed as $\A^{(0)}$ plus
the transversal gauge for $\curl\A^\infty$, but the derivatives of the
latter need not decay integrably.

Lemma~\ref{Lh} with $m_0=1/2+3\delta$,
$m_1=3/2+2\delta$ and $\Delta=1/2+\delta$ yields a decomposition
$\A^\infty=\A^{(1)}+\A^{(2)}$, such that $\A^{(1)}$ is short-range and
$\partial^\gamma A_k^{(2)}(\x)=\O(|\x|^{-m_j'})$ for $|\gamma|=j$,
with $m_1'=m_1=3/2+2\delta$ and $m_2'=2+3\delta$. Now define
$\A^s:=\A^{(0)}+\A^{(1)}=\A^c-\A^{(2)}$, then $\A^s$ is
continuous, weakly differentiable, and short-range.
The longitudinal part of $\A^{(2)}$ need not decay integrably. Consider the
decomposition $\B=\B^s+\B^r$ with $\B^s=\curl\A^s$ in $\S'$ and
$\B^r=\curl\A^{(2)}$ in $C^\infty$, which satisfies $\div\B^s=\div\B^r=0$.
We have $\B^r(\x)=\O(|\x|^{-3/2-2\delta})$ and
$\partial_i B_k^r(\x)=\O(|\x|^{-2-3\delta})$. Define $\A^r$ as the
transversal gauge vector potential belonging to $\B^r$, then
$\A^r\in C^\infty$ with $\A^r(\x)=\O(|\x|^{-1/2-2\delta})$. By
differentiating (\ref{tg}) under the integral and an analogous estimate,
$\partial_i A_k^r(\x)=\O(|\x|^{-1-3\delta})$ is obtained.
Now $\A:=\A^s+\A^r$ yields the desired gauge. Note that
 $\div\A=\div\A^r-\div\A^{(2)}\in C^\infty$ decays integrably. \mybox

\section{Inversion of X-Ray Transforms} \label{Sx}
From the asymptotics of the scattering operator $S$, we will know the line
integral of $\A$ along all straight lines in $\R^\nu$,
up to adding a function of the direction $\bomega\in S^{\nu-1}$:
 \be \label{x1} a(\bomega,\,\x):=\int_{-\infty}^\infty
 \bomega\cdot\A(\x+\bomega t)\,dt \ . \ee

\begin{prop}[Inversion of X-Ray Transforms] \label{Px}
Suppose $\A$ is an unknown vector potential of medium range, and
the line integral \emph{(\ref{x1})} is given for all $\x\in\R^\nu$ and
 $\bomega\in S^{\nu-1}$, up to adding a function $f(\bomega)$. Then

$1$. The distribution $\B:=\curl\A\in\S'$ is determined uniquely.

$2$. Assume in addition that $\B$ is a magnetic field of medium decay. On
a.e.~plane, the X-ray transform of the normal component of $\B$ is obtained
from \emph{(\ref{x0})} below.
\end{prop}

Under stronger decay assumptions on $\A$, item~1 is due to
\cite{wjdip, wjde}, and under stronger regularity assumptions, item~2
is found, e.g., in \cite{itd, nab}.

\textbf{Proof}: 1.
Given $\phi\in\S(\R^\nu,\,\R^{\nu'})$, we will need a vector field
 $\psi\in C^\infty(\R^\nu,\,\R^\nu)$ with
 \be \label{x3} \curl\phi(z) = 2\int_{\R^\nu}
 \dfrac{\y\,\y\cdot\psi(\z-\y)}{|\y|^{\nu+1}} \,dy \ . \ee
By Fourier transformation, this equation is equivalent to
 \be \i\p\times\hat\phi(\p)=c_\nu\,\dfrac1{|\p|^3}\,
 \Big(|\p|^2\hat\psi(\p)-\p\,\p\cdot\hat\psi(\p)\Big) \ee
for some $c_\nu>0$. Choose the following solution $\psi\in\C^\infty$:
 \ban \hat\psi(\p) &:=& \i c_\nu^{-1}|\p|\,\p\times\hat\phi(\p) \nonumber
 \;=\; c_\nu^{-1}|\p|^{-1}\,\Big(|\p|^2\,\i\p\times\hat\phi(\p)\Big) \ ,
 \\[1mm]  \psi(x) &=& -c_\nu' \int_{\R^\nu} \label{x4}
 |\x-\u|^{-(\nu-1)}\,\Delta\curl\phi(\u)\,du \ . \ean
To determine the decay properties of $\psi$, split
$|\x|^{-(\nu-1)}=f_1(\x)+f_2(\x)$ such that $f_1$ has compact support and
 $f_2\in C^\infty$, and split $\psi=\psi_1+\psi_2$ according to (\ref{x4}).
Integrating by parts three times shows that $\psi_2$ is the convolution of
$\phi$ with a $C^\infty$-kernel, which is $\O(|\x|^{-(\nu+2)})$, thus in
$L^1$. Therefore $\psi\in L^1$. Now suppose that
 $a(\bomega,\,\x)=\int_{-\infty}^\infty \bomega\cdot\A(\x+\bomega t)\,dt$
is known up to a constant depending on $\bomega$, then
 \be \int_{\R^\nu} a(\bomega,\,\x)\,\bomega\cdot\psi(\x)\,dx=
 \int_{-\infty}^\infty \int_{\R^\nu} \bomega\cdot\A(\x+\bomega t)
 \,\bomega\cdot\psi(\x)\,dx\,dt \ee
is known: the unknown constant is canceled, since $\hat\psi(0)=0$ gives
 $\int_{\R^\nu}\psi\,dx=0$. Consider polar coordinates $\y=\bomega t$,
$dy=|\y|^{\nu-1}d\omega\,dt$ to obtain
 \ba  \int_{S^{\nu-1}}\int_{\R^\nu}
 a(\bomega,\,\x)\,\bomega\cdot\psi(\x)\,dx\,d\omega
 &=& 2\int_{\R^\nu}\int_{\R^\nu} \dfrac{\y\cdot\A(\x+\y)
 \,\y\cdot\psi(\x)}{|\y|^{\nu+1}}\,dx\,dy \\[1mm]
 =\, 2\int_{\R^\nu}\int_{\R^\nu} \dfrac{\y\cdot\A(\z)
 \,\y\cdot\psi(\z-\y)}{|\y|^{\nu+1}}\,dy\,dz
 &=& \int_{\R^\nu}\A(\z)\cdot\curl\phi(\z)\,dz \ea
by (\ref{x3}), thus the distribution $\B=\curl\A\in\S'$ has been computed.

2. Fix unit vectors $\bomega,\,\tilde\bomega\in S^{\nu-1}$. If $\A\in C^1$,
thus $\B$ is continuous, we have
 \be \label{x2} \dfrac{\partial}{\partial u}
 \int_{-\infty}^\infty \bomega\cdot\A (\x+\tilde\bomega u+\bomega t) \,dt
 = \int_{-\infty}^\infty (\tilde\bomega\times\bomega)\cdot\B
 (\x+\tilde\bomega u+\bomega t) \,dt \ . \ee
(Proof by Stokes' Theorem, or with
 $(\tilde\bomega\times\bomega)\cdot(\nabla\times\A)=
 (\tilde\bomega\cdot\nabla)(\bomega\cdot\A)-
 (\bomega\cdot\nabla)(\tilde\bomega\cdot\A)$.) Thus the X-ray transform of
the component of $\B$ in the direction of $\tilde\bomega\times\bomega$ is
obtained on every plane normal to $\tilde\bomega\times\bomega$. (It is
natural, but not required, to assume $\tilde\bomega\cdot\bomega=0$.) The
left hand side of (\ref{x2}) does not depend on the gauge of $\A$, since
$\partial_u f(\bomega)=0$. Now $\B\in L^p$, and $\A$ is of medium range. Then
(\ref{x2}) remains true for a.e.~$\x\in\R^\nu$ and a.e.~$u\in\R$. The proof
is given for $\nu=3$, $\bomega=(0,\,1,\,0)^\tr$ and
$\tilde\bomega=(1,\,0,\,0)^\tr$: We have
 \be \int_{-\infty}^\infty A_2(u,\,t,\,x_3) \,dt
 -\int_{-\infty}^\infty A_2(0,\,t,\,x_3) \,dt
 = \int_0^u\int_{-\infty}^\infty B_3(v,\,t,\,x_3)\,dt\,dv \ee
for every $u$ and almost every $x_3\,$, by integrating with respect to
$x_3\,$, an approximation argument, and Fubini. For almost every $x_3\,$,
both sides are well-defined, and the right hand side is weakly differentiable
with respect to $u$. (\ref{x2}) may be rewritten as
 \be \label{x0} (\tilde\bomega\cdot\nabla)\,a(\bomega,\,\x)
 = \int_{-\infty}^\infty (\tilde\bomega\times\bomega)\cdot\B
 (\x+\bomega t) \,dt \ , \ee
where $\tilde\bomega\cdot\nabla$ denotes a weak directional derivative,
which exists for a.e.~$\x$. \mybox

$\B$ is reconstructed from the X-ray transform according to
 \cite{he, ew2, wjde}. We have considered the normal component of $\B$
on almost every plane:

\begin{rmk}[Trace of $\B$ on a Plane] 
Suppose $\B:\R^3\to\R^3$ is a magnetic field of medium decay. If $\B$ is
continuous except for a jump discontinuity transversal to a surface, the
condition $\div\B=0$ in $\S'$ implies that the normal component of $\B$
is continuous. If $\B\in L^p$ with $p>3$, I do not know if there is a
kind of trace operator, which defines the restriction of $\bomega\cdot\B$
to every plane normal to $\bomega$, in $L_\loc^1(\R^2)$ or in
$L^2(\R^2)$. (This restriction is well-defined as a distribution in
$\S'(\R^2,\,\R)$, by employing a vector potential.)
\end{rmk}

\section{The Direct Problem of Scattering Theory} \label{Sgo}
Now vector potentials of medium range are applied to a nonrelativistic
scattering problem: this section contains the proof of
Thm.~\ref{The}, item~1.

\subsection{Definition of Hamiltonians}
Our Hilbert space is $\H=L^2(\R^\nu,\,\C)$ for the Schr\"odinger equation,
and $\H=L^2(\R^\nu,\,\C^2)$ for the Pauli equation. In the latter case, the
Pauli matrices $\sigma_i\in\C^{2\times2}$ are employed \cite{th}. The free
time evolution is generated by the free Hamiltonian
$H_0=-\frac1{2m}\Delta=\frac1{2m}\p^2$. It is self-adjoint with domain
$D_{H_0}=W^2$, a Sobolev space. In an external electromagnetic field, the
Pauli Hamiltonian is defined formally by the following expressions:
 \ban H &=& \label{ham1}
 \frac1{2m}\,\Big[(\p-\A)^2 -\bsigma\cdot\B\Big] + A_0 \\[1mm]
 &=& \frac1{2m}\,\Big[\p^2-2\A\cdot\p +  \label{ham2}
 \i\div\A+\A^2-\bsigma\cdot\B\Big] + A_0 \ . \ean
Domains will be specified below by employing perturbation theory of operators
and quadratic forms, cf.~\cite{rs2, rs1}. The following quadratic form
is needed as well:
 \be q_\A(\psi,\,\psi) :=  \label{ham3}
 \frac1{2m}\,\Big\|(\p-\A)\psi\Big\|^2
 +\Big(\psi,\,(-\frac1{2m}\bsigma\cdot\B+A_0)\psi\Big) \ . \ee

\begin{lem}[Pauli- and Schr\"odinger Operators]  \label{Lps}
Suppose $\B$ is a magnetic field of medium decay and $\A$ is a corresponding
vector potential of medium range, and $A_0$ is a scalar potential of short
range, cf.~\emph{(\ref{vs})}. The Pauli operator $H$ is defined in item~$1$:

$1$. There is a unique self-adjoint operator $H$, such that its form domain
is the Sobolev space $W^1$, and the quadratic form corresponding to $H$
equals $q_\A$ according to \emph{(\ref{ham3})}.

$2$. Suppose $\A'=\A+\grad\lambda$ is of medium range as well,
and define $H'$ in terms of $q_{\A'}$ analogously to item~$1$. Then
$D_{H'}=\e{\i\lambda(\x)}D_H$ and
$H'=\e{\i\lambda(\x)}H\e{-\i\lambda(\x)}$.

$3$. Suppose the distribution $\div\A\in\S'$ is a bounded function. Then
$D_H=W^2$, and $H$ satisfies \emph{(\ref{ham2})}.

For the Schr\"odinger operator, the term $-\bsigma\cdot\B$ is omitted, and
$\curl\A\in\S'$ need not be a function.
\end{lem}

\textbf{Proof}: 1. The quadratic form (\ref{ham3}) is well-defined on $W^1$,
which is the form domain of $H_0\,$. It satisfies
 $|q_\A(\psi,\,\psi)|\le a(\psi,\,H_0\psi)+\O(\|\psi\|^2)$ for some $a<1$:
the bound for $A_0$ is $<1$ since $A_0$ is short-range, and the bounds of
the other terms are arbitrarily small. Now $H$ is obtained from the
KLMN Theorem.

2. $\grad\lambda=\A'-\A$ is bounded and continuous. The mapping
$\psi\mapsto\e{\i\lambda}\psi$ is unitary in $\H$ and sending $W^1$ to
itself. On $W^1$ we have
 \be \e{\i\lambda(\x)}\Big(\p-\A(\x)\Big)\,\e{-\i\lambda(\x)}
 = \p-\grad\lambda(\x)-\A(\x)=\p-\A'(\x) \ , \ee
thus $q_{\A'}(\psi,\,\psi)=q_\A(\e{-\i\lambda}\psi,\,\e{-\i\lambda}\psi)$
for $\psi\in W^1$. Since $H$ and $H'$ are determined uniquely by the
quadratic forms, we have $H'=\e{\i\lambda}H\e{-\i\lambda}$.

3. Now a sum of distributions is a bounded function. Denoting it by
$\div\A$, we have
 $\int_{\R^\nu} \A\cdot\grad\phi +(\div\A)\phi\,dx=0$ for $\phi\in\S$.
(If the distributions $\curl\A$ and $\div\A$ are in $L_\loc^2\,$, then all
$\partial_iA_k\in L_\loc^2$ by elliptic regularity. But here we include the
more general case of a Schr\"odinger operator as well, where $\div\A$ is a
function but $\curl\A$ is not.) The expression (\ref{ham2}) is a symmetric
operator on $\S$. By the Kato--Rellich Theorem, its closure is a self-adjoint
operator $\tilde H$ with $D_{\tilde H}=D_{H_0}=W^2$. We compute
 $q_\A(\psi,\,\psi)=(\psi,\,\tilde H\psi)$ for $\psi\in\S$. By the KLMN
Theorem, $\S$ is a form core for $H$, thus $H=\tilde H$. \mybox

\subsection{Existence of Wave Operators}  \label{subex}
Consider a Pauli operator $H$ according to Lemma~\ref{Lps}. We assume that
$\A$ is given by item~4 of Thm.~\ref{Tg}, i.e., $\A=\A^s+\A^r$ with $\A^s$
short-range, $\A^r$ transversal, and $\div\A$
short-range. Thus $H$ is given by (\ref{ham2}). Existence of the wave
operators will be shown by estimating the Cook integral
 \be \label{cook} \Omega_\pm\,\psi :=
 \lim_{t\to\pm\infty} \e{\i Ht}\,\e{-\i H_0t}\,\psi = \psi +
 \i\int_0^{\pm\infty} \e{\i Ht}\,(H-H_0)\,\e{-\i H_0t}\,\psi\,dt \ . \ee
The integral is well-defined on a finite $t$-interval if
 $\psi\in D_{H_0}=D_H=W^2$. We have
 \be \label{arvs} H-H_0 =
 -\frac1m\,\A^r\cdot\p + V^s \ ,\quad V^s:= \frac1{2m}\,
 \Big\{-2\A^s\cdot\p+\i\div\A+\A^2-\bsigma\cdot\B\Big\}+A_0 \ .\ee
Fix $\psi\in\S$ with $\hat\psi\in C_0^\infty(\R^\nu\setminus\{0\})$ and
$\eps>0$ with $\hat\psi(\p)\equiv0$ for $|\p|<\eps m$.
Choose $g\in C_0^\infty(\R^\nu,\,\R)$ with $g(\p)\equiv1$ on
$\supp(\hat\psi)$ and consider the decomposition
\ban \Big\|A_0\,\e{-\i H_0t}\,\psi\Big\| \label{dece1}
 &\le& \Big\|A_0\,g(\p)\,F(|\x|\ge\eps|t|)\Big\|\cdot
  \Big\|\e{-\i H_0t}\,\psi\Big\| \\[1mm]
 &+& \Big\|A_0\,g(\p)\Big\|\cdot \label{dece2}
  \Big\|F(|\x|\le\eps|t|)\,\e{-\i H_0t}\,\psi\Big\| \ . \ean
The first summand has an integrable bound since $A_0$ is a short-range
potential (in (\ref{vs}), the resolvent may be replaced with $g(\p)$).
The second term can be estimated by any inverse power of $|t|\,$,
by a standard nonstationary phase estimate for propagation into the
classically forbidden region \cite{th}, since the speed is bounded below
by $\eps$. The remaining terms in $V^s$ are treated
analogously, observing the decay properties of $\A^s$, $\A^2$, $\div\A$, and
$\B$. In the term $\A^s\cdot\p$ the space decomposition is introduced between
$\A^s$ and $\p$. The term $\A^r\cdot\p$ in (\ref{arvs}) is controlled with
the technique of \cite{ltn, th}: write $\A^r(\x)=-\x\times\G(\x)$ and note
that $(\G\times\x)\cdot\p=\G\cdot\mathbf{L}$ with the angular momentum
$\mathbf{L}=\x\times\p$. Now $\mathbf{L}$ is commuting with $H_0\,$, and
$\G$ is a short-range term, thus we have obtained an integrable bound $h(t)$
for the integrand in (\ref{cook}). The integral exists as a Bochner integral
or as an improper Riemann integral. Thus the limit exists for a dense set of
states $\psi$, and the wave operators exist as a strong limit on $\H$.
(We have not used the fact that $\partial_iA_k$ decays integrably, but it
will be needed in the relativistic case \cite{wjm2}.) In an
arbitrary gauge $\A'$ for the given $\B$, existence of the wave operators
follows now from the transformation formula (\ref{gtpr}). The scattering
operator is defined by $S:=\Omega_+^*\,\Omega_-\,$. The Schr\"odinger
equation is treated analogously, by omitting the term $-\bsigma\cdot\B$.
(Existence of $\Omega_\pm$ can be shown for every medium-range $\A$, without
any assumptions on $\div\A$ or $\curl\A$, by quadratic form techniques. This
proof requires an additional regularization, and it is not suitable
for obtaining a high-energy limit.)

\subsection{Gauge Transformation}
Suppose $\A$ and $\A'$ are vector potentials of medium range with
$\curl\A=\curl\A'$ in $\S'$, thus $\A'=\A+\grad\lambda$ and
$\Lambda(\x)=\lim_{r\to\infty}\lambda(r\x)$ is continuous on
$\R^\nu\setminus\{0\}$ by Thm.~\ref{Tg}. We claim
 \be \label{cllam} \slim_{t\to\pm\infty}
 \e{\i H_0t}\lambda(\x)\,\e{-\i H_0t} = \Lambda(\pm\p) \ . \ee
Note that $\Big(\lambda(\x)-\Lambda(\x)\Big)\,(H_0+\i)^{-1}$
is compact, thus
 \be \slim_{t\to\pm\infty}
 \e{\i H_0t}\,\Big(\lambda(\x)-\Lambda(\x)\Big)\,\e{-\i H_0t} = 0 \ . \ee
Moreover, since $\Lambda$ is 0-homogeneous, we may multiply the argument
with $\pm m/t>0$:
 \ban \e{\i H_0t}\Lambda(\x)\,\e{-\i H_0t}
 &=& \Lambda(\x+t\p/m) \;=\; \Lambda(\pm m\x/t\pm\p) \\[1mm]
 &=& \e{-\i m\x^2/(2t)}\Lambda(\pm\p)\,\e{\i m\x^2/(2t)}
 \;\to\; \Lambda(\pm\p) \label{gtprf} \ean
strongly as $t\to\pm\infty$, since $\x^2/t\to0$ pointwise for $\x\in\R^\nu$.
This proves (\ref{cllam}). Now consider the
Hamiltonians $H$ and $H'=\e{\i\lambda(\x)}H\e{-\i\lambda(\x)}$ according to
Lemma~\ref{Lps}, and suppose that $\Omega_\pm$ exist. Then the wave operators
 \ban \Omega_\pm' \;:=\; \slim_{t\to\pm\infty}\e{\i H't}\e{-\i H_0t}
 &=& \slim_{t\to\pm\infty}\e{\i\lambda(\x)}\Big(\e{\i Ht}\e{-\i H_0t}\Big)\,
 \Big(\e{\i H_0t} \e{-\i\lambda(\x)} \e{-\i H_0t}\Big) \nonumber \\[1mm]
 &=& \e{\i\lambda(\x)}\Omega_\pm\,\e{-\i\Lambda(\pm\p)} \label{gtpr} \ean
exist as well, and the scattering operators satisfy
 $S'=\e{\i\Lambda(\p)}S\,\e{-\i\Lambda(-\p)}$. The gauge transformation
formula is employed, e.g., in \cite{rab, tss, wab, ycd}. 
The proof (\ref{gtprf}) seems to be new.
The analogous formula for the Dirac equation is found in
\cite{wjdip, wjde, wjm2}. If $\A-\A'$ is short-range,
then $\Lambda$ is constant, and $S'=S$.

\subsection{Asymptotic Completeness of Wave Operators}
The wave operators $\Omega_\pm$ are called asymptotically complete, if
every ``scattering state'' in the continuous subspace of $H$ is asymptotic to
a free state, i.e., $\Ran(\Omega_-)=\Ran(\Omega_+)=\H^{cont}(H)=\H^{ac}(H)$,
which implies that $S$ is unitary. Consider again the special gauge
$\A=\A^s+\A^r$: In the case of $\A^s=0$ and $A_0=0$, completeness was shown
in \cite{ltn, elr} by the Enss geometric method. The proof shall extend to
our case, since the additional short-range terms
can be included with standard techniques, but I have not checked the details.
In an arbitrary gauge $\A'=\A+\grad\lambda$, completeness is carried over by
the gauge transformation (\ref{gtpr}): we have
 $\Ran(\Omega_\pm')=\e{\i\lambda(\x)}\Ran(\Omega_\pm)
 =\e{\i\lambda(\x)}\H^{ac}(H)=\H^{ac}(H')$.
Under the stronger assumptions
$\B\in L_\loc^4$ and $\curl\B(\x)=\O(|\x|^{-(2+\delta)})$, completeness was
shown by Arians \cite{ardip} in the transversal gauge. Her proof employs
a phase space cutoff of the form $f(\p-\A(\x))$, which can be defined by
a Fourier transform.

\subsection{Modified Wave Operators} \label{subik}
For $\A$ of medium decay, the unmodified wave operators exist although
$\A(\x)$ need not decay integrably. We shall compare them to modified wave
operators: These exist when $\A(\x)=\A^s(\x)+\A^l(\x)$, where $\A^s$ is
short-range, and $\A^l$ is $C^\infty$, with $\A^l(\x)=\O(|\x|^{-\delta})$,
and with decay assumptions on its derivatives. The Dollard wave operators
are obtained from a time-dependent modification. If $\A$ is of medium
range, it could be done in the form
 \be \label{omd} \Omega_\pm^D:=\slim_{t\to\pm\infty} \e{\i Ht}\,U^D(t)\ ,
 \qquad U^D(t):=
 \exp\Big\{-\i H_0t+\i\int_0^t\p\cdot\A^l(s\p)\,ds\Big\} \ , \ee
 thus $\Omega_\pm^D=\Omega_\pm\,\exp
 \Big\{\i\int_0^{\pm\infty}\p\cdot\A^l(s\p)\,ds\Big\}$. By choosing $\A^l$
transversal, the modification is vanishing. The time-independent modification
of Isozaki--Kitada is employed, e.g., in \cite{tss, nab, ryab, yhs}.
With Fourier integral operators $J_\pm$ we have
 \be \label{omj} \Omega_\pm^J:=\slim_{t\to\pm\infty}
 \e{\i Ht}\,J_\pm\,\e{-\i H_0t} \ , \qquad
 J_\pm \,:\,\e{\i\q\x} \mapsto u_\pm^\q(\x) \ee
for smooth $\A(\x)=\O(|\x|^{-\delta})$. Here $u_\pm^\q(\x)$ is an
approximate generalized eigenfunction of $H$ with incoming or outgoing
momentum $\q$, e.g., according to \cite{yhs}:
 \be \label{uj} u_\pm^\q(\x) := \exp\Big\{\,\i\q\cdot\x+\i\int_0^{\pm\infty}
 \bomega\cdot\A(\bomega s)-\bomega\cdot\A(\x+\bomega s)\,ds\,\Big\} \ ,
 \quad \q=|\q|\bomega \ . \ee
Under a change of gauge, $\A\to\A'=\A+\grad\lambda$ with
$\grad\lambda(\x)=\O(|\x|^{-\delta})$, the FIOs and the
modified wave operators are transformed according to
 \be \label{gtoj} J_\pm' = \e{\i\lambda(\x)-\i\lambda(0)}\,J_\pm \ , \qquad
 {\Omega_\pm^J}' = \e{\i\lambda(\x)-\i\lambda(0)}\,\Omega_\pm^J \ , \ee
and the modified scattering operator $S^J:={\Omega_+^J}^*\,\Omega_-^J$ is
gauge-invariant. (This is not the case when $\bomega\cdot\A(\bomega s)$ is
omitted from the integrand in (\ref{uj}), or in the idealized
Aharanov--Bohm experiment, where $\A(\x)$ is unbounded at $\x=0$
and $\lambda(\x)\equiv\Lambda(\x)$ \cite{nab, wab, ryab}. Then
$\Omega_\pm^J=\Omega_\pm$ and $S^J=S$ transform according
to (\ref{gts}).) When $\A$ is smooth and medium-range,
\cite[Lemma~2.2]{nfo} implies
 \be \label{sjs} S^J = \e{-\i a(\p)}\,S\,\e{\i a(-\p)} \ , \qquad
 a(\p) := \int_0^\infty\p\cdot\A(s\p)\,ds \ . \ee
This may be taken as the definition of $S^J$ when $\A$ is not smooth.
Cf.~Sec.~\ref{subscs}.

\section{High-Energy Limit and Inverse Scattering} \label{Si}
Consider the scattering of a state
$\e{\i\u\x}\psi$, where $\u=u\bomega$. The position operator $\x$ is
generating a translation by $\u$ in momentum space, and the limit of
$\e{-\i\u\x}\Big(S\,\e{\i\u\x}\psi\Big)$ gives the high-energy
asymptotics of the scattering process, as $u\to\infty$ for a fixed
direction $\bomega\in S^{\nu-1}$.

\subsection[High-Energy Limit of S]{High-Energy Limit of $S$} \label{subhe}
Applying the momentum-space translation by $\u=u\bomega$ to the free
Hamiltonian gives
 \be \e{-\i\u\x}H_0\,\e{\i\u\x} =\frac1{2m}(\p+\u)^2
 = \frac1{2m}u^2 + \frac um\Big(\bomega\cdot\p + \frac mu H_0\Big) \ . \ee
In the free time evolution, the first term is a rapidly oscillating phase
factor, which cancels with the corresponding term for $H$ in the Cook
integral (\ref{cooku}). Rescaling the time $t'=ut/m$, the free time
evolution is generated by $\bomega\cdot\p + \frac m u H_0\to\bomega\cdot\p$
as $u\to\infty$. For the Hamiltonian $H$, note that $\A\cdot\p$ becomes
$\A\cdot(\p+\u)$ before rescaling, and $H$ is replaced with
$H_\u:=\bomega\cdot(\p-\A) + \frac m u H$. Employing the special gauge
$\A=\A^s+\A^r$ according to item~4 of Thm~\ref{Tg}, consider
 \ban \label{cooku} && \e{-\i\u\x}\,\Omega_\pm\,\e{\i\u\x}\,\psi \\[1mm]
 &=& \psi + \i\int_0^{\pm\infty} \e{\i H_\u t} \nonumber
 \,\Big\{-\bomega\cdot\A+\frac mu(H-H_0)\Big\}\,
 \e{-\i(\bomega\p+{\scriptstyle\frac mu}H_0)t}\,\psi\,dt \ean
Assume $\psi\in\S$ with $\hat\psi\in C_0^\infty$. The velocity operator
corresponding to the translated and rescaled time evolution is
$\bomega+\p/u$. Fix $0<\eps<1$ and $u_0>0$, such that $\supp\hat\psi$
is contained in the ball $|\p|\le u_0(1-\eps)$, then the speed is bounded
below by $\eps$ for $u\ge u_0\,$. By the standard techniques from
Sec.~\ref{subex}, i.e., the
decomposition (\ref{dece1})--(\ref{dece2}), an integrable bound $h(t)$ is
obtained for the integrand in (\ref{cooku}), uniformly for $u\ge u_0\,$.
The critical term is
$-\A^r\cdot(\bomega+\p/u)=-\G\cdot[\x\times(\bomega+\p/u)]$ with
$\A^r=\G\times\x$. Again, the
translated angular momentum $\x\times(\bomega+\p/u)$ is commuting with
the translated free time evolution, and $\G(\x)$ is short-range. By the
Dominated Convergence Theorem (for the $\H$-valued Bochner integral),
the limit $u\to\infty$ is interchanged with the integration:
 \ban \lim_{u\to\infty} \e{-\i\u\x}\,\Omega_\pm\,\e{\i\u\x}\,\psi
 &=& \psi + \i\int_0^{\pm\infty} \e{\i\bomega(\p-\A)t}
 \,(-\bomega\cdot\A)\,\e{-\i\bomega\p t}\,\psi\,dt \\[1mm]
 &=& \lim_{t\to\pm\infty} \e{\i\bomega(\p-\A)t}\,\e{-\i\bomega\p t}
  \,\psi\\[1mm]
 &=& \exp\Big\{ -\i\int_0^{\pm\infty}
 \bomega\cdot\A(\x+\bomega t)\,dt \Big\} \,\psi \ . \ean
We have employed the fact that $H_\u\to\bomega\cdot(\p-\A)$ in the strong
resolvent sense. The last step is verified from a differential equation
\cite{ar1, wjde}. A density argument yields strong convergence of
$\Omega_\pm\,$, and (\ref{hes}) follows from
$S=\Omega_+^*\,\Omega_-$ and the strong convergence of $\Omega_+^*\,$.
(If $\Omega_u$ are isometric, $\Omega_\infty$ is unitary, and
$\Omega_u\to\Omega_\infty$ strongly, then $\Omega_u^*\to\Omega_\infty^*$
strongly because $\Omega_u^*-\Omega_\infty^*
=\Omega_u^*(\Omega_\infty-\Omega_u)\Omega_\infty^*\,$.)
In an arbitrary medium-range gauge $\A'$
consider $\A'=\A+\grad\lambda\,$, the limit $\Lambda$ according to
Thm.~\ref{Tg}, and the transformation formula (\ref{gts}). We obtain
\ba \slim_{u\to\infty} \,\e{-\i\u\x} \,S'\, \e{\i\u\x} &=&
 \e{\i\Lambda(\bomega)}\,
 \exp\Big\{ \i\int_{-\infty}^\infty \bomega\cdot\A (\x+\bomega t)
 \,dt \Big\} \, \e{-\i\Lambda(-\bomega)} \\[1mm]
 &=& \exp\Big\{ \i\int_{-\infty}^\infty \bomega\cdot\A' (\x+\bomega t)
 \,dt \Big\} \ , \ea
since $\lambda(\x+\bomega t)\to\Lambda(\pm\bomega)$ pointwise for
$t\to\pm\infty$. Thus the high-energy limit (\ref{hes}) is established
for an arbitrary gauge $\A$. Under stronger decay assumptions,
error bounds are given in Cor.~\ref{Ce}. The same proof applies to the
Schr\"odinger equation by omitting the term $-\bsigma\cdot\B$.

The simplification of the scattering process to a mere phase change at
high energies has a geometric interpretation, cf.~\cite{ew1, eji}: For a
state $\psi$ with momentum support in a ball of radius $mR$ around
$\u=u\bomega$ and large $u=|\u|$, translation dominates over spreading
of the wave packet in the free time evolution. On the physical time scale
$t$, the region of strong interaction is traveled in a time $t\simeq m/u$,
and the effective diameter of the wave packet is increased by
$\approx Rt\simeq1/u$.

\subsection[Reconstruction of B]{Reconstruction of $\B$}
In the inverse scattering problem, $A_0\,$, $\B$ and $\A$ are unknown, and
the high-energy limit of $S$ is known up to a gauge transformation. The
absolute phase of (\ref{hes}) is not gauge-invariant, but we assume that the
relative phase is observable. The exponential function is $2\pi\i$-periodic,
but by the continuity of $\A$, the integral transform of $\A$ is obtained up
to a direction-dependent constant, and the magnetic field $\B$ is
reconstructed according to Prop~\ref{Px}. This concludes the proof of
Thm.~\ref{The} for the Pauli operator and the Schr\"odinger operator.

\subsection[Reconstruction of A\_0]{Reconstruction of $A_0$} \label{subrec}
For the Schr\"odinger equation with short-range electromagnetic fields,
Arians \cite{ar1} first reconstructs $\B$ from (\ref{hes}), and then she
reconstructs $A_0$ from the $1/u$-term of the high-energy asymptotics. The
following theorem is quite similar, but the proof will be modified. Similar
results for $\B\in C^\infty$ are obtained by Nicoleau in \cite{nfo} using
Fourier integral operators.

\begin{thm}[Reconstruction of $A_0$ (Arians)] \label{Ta}
Consider a short-range electrostatic potential $A_0$ according to
\emph{(\ref{vs})}, and a magnetic field $\B:\R^3\to\R^3$
or $\B:\R^2\to\R$. Suppose $\B$ and $\curl\B$ are continuous and both decay
as $\O(|\x|^{-\mu})$ with $\mu>2$, and that the flux of $\B$ is vanishing
if $\nu=2$. $\A$ is a corresponding vector
potential of short range. Set
 $a(\bomega,\,\x):=\int_{-\infty}^\infty\bomega\cdot\A(\x+\bomega t)\,dt$.
The scattering operator $S$ for the Schr\"odinger- or Pauli equations has
the following high-energy asymptotics:
 \ban \lim_{u\to\infty}\, \dfrac{u}{m}\,\e{-\i\u\x} \,\Big(
 S-\e{\i a}\Big)\, \e{\i\u\x} \,\psi  \nonumber
 &=& -\i\,\e{\i a}\int_{-\infty}^\infty
  A_0(\x+\bomega t)\,\psi \,dt \\[1mm]
 && -\i\,\e{\i a}\int_{-\infty}^0  \label{a1}
  K_-^\bomega(\x+\bomega t,\,\p)\,\psi \,dt \\[1mm] \nonumber
 && -\i\int_0^\infty
 K_+^\bomega(\x+\bomega t,\,\p)\,\e{\i a}\,\psi \,dt \ean
for $\hat\psi\in C_0^\infty$. The operators $K_\pm^\bomega(\x,\,\p)$ are
defined in \emph{(\ref{a6})}, they depend on $\B$ but not on $A_0$.
Thus $\B$ and $A_0$ are reconstructed uniquely from $S$.
\end{thm}

\textbf{Proof}: For every $\bomega\in S^{\nu-1}$ and the signs $\pm$,
consider the vector potentials
 \ban \A_\pm^\bomega(\x) \;:=\; \label{a2}
 \int_0^{\pm\infty}\bomega\times\B(\x+\bomega s)\,ds
 &=& \A(\x)+\grad\lambda_\pm^\bomega(\x) \\[1mm]
 \mbox{with}\qquad \lambda_\pm^\bomega(\x) &:=&
 \int_0^{\pm\infty}\bomega\cdot\A(\x+\bomega s)\,ds \ . \ean
This definition is motivated by \cite{tss, ar1, nfo}, and the transformation
(\ref{a2}) is verified with $\bomega\times(\nabla\times\A)=
 \nabla(\bomega\cdot\A)-(\bomega\cdot\nabla)\A$. $\A_\pm^\bomega$ and
$\grad\lambda_\pm^\bomega$ are bounded and continuous. $\A_\pm^\bomega(\x)$
decays as $|\x|^{-(\mu-1)}$ in the half-space $\bomega\cdot\x\to\pm\infty$,
but in general it does not decay for $\bomega\cdot\x\to\mp\infty$, thus
$\A_\pm^\bomega$ is not of medium range. We have
 $\curl\A_\pm^\bomega=\B$ and $\div\A_\pm^\bomega(\x)=
 -\int_0^{\pm\infty}\bomega\cdot\curl\B(\x+\bomega s)\,ds$ in $\S'$. The
latter function is bounded, continuous, and decays as $|\x|^{-(\mu-1)}$ for
$\bomega\cdot\x\to\pm\infty$. The Hamiltonian $H_\pm^\bomega$ is defined by
(\ref{ham2}) with $\A_\pm^\bomega$ instead of $\A$, it satisfies
 $H_\pm^\bomega=\e{\i\lambda_\pm^\bomega(\x)}\,H\,
  \e{-\i\lambda_\pm^\bomega(\x)}$.
Analogously to the gauge transformation formula (\ref{gtpr}) we have
 \be \label{a3} \Omega_\pm\,\psi =
 \e{-\i\lambda_\pm^\bomega(\x)} \lim_{t\to\pm\infty}
 \e{\i H_\pm^\bomega t}\,\e{-\i H_0t}\,\psi \ , \ee
when the support of $\hat\psi$ is contained in the half-space
$\pm\bomega\cdot\p>0$. To verify that the analog of
$\Lambda(\pm\p)$ is vanishing, a nonstationary phase estimate
is employed for a dense set of states. For $\hat\psi\in C_0^\infty$,
consider the translated and rescaled Cook integral
 \ban \label{cooka} && \frac um\,\e{-\i\u\x}\,
 \Big(\e{\i\lambda_\pm^\bomega(\x)}\,
 \Omega_\pm-1\Big)\,\e{\i\u\x}\,\psi \\[1mm]
 &=& \nonumber
 \i\int_0^{\pm\infty} \e{\i(\bomega\p+{\scriptstyle\frac mu}H_\pm^\bomega)t}
 \,\Big\{A_0(\x)+K_\pm^\bomega(\x,\,\p)\Big\}\,
 \e{-\i(\bomega\p+{\scriptstyle\frac mu}H_0)t}\,\psi\,dt \ean
with the symmetric operators (omit $-\bsigma\cdot\B$ in the
Schr\"odinger case)
 \be \label{a6} K_\pm^\bomega(\x,\,\p) :=
 \frac1{2m}\,\Big(-2\A_\pm^\bomega(\x)\cdot\p+\i\div\A_\pm^\bomega(\x)
  +(\A_\pm^\bomega(\x))^2-\bsigma\cdot\B(\x)\Big) \ . \ee
Note that $\bomega\cdot\A_\pm^\bomega=0$, thus the main contribution from
(\ref{cooku}) is missing, and a common factor $1/u$ was extracted from the
remaining terms. By the same uniform estimates as in Sec.~\ref{subhe}, the
limit $u\to\infty$ can be performed under the integral. Since the translated
and rescaled generators of the time evolutions are converging to
$\bomega\cdot\p$ in the strong resolvent sense, (\ref{cooka})
is converging to
 \be \label{a5}  \i\int_0^{\pm\infty} \,\Big\{A_0(\x+\bomega t)
 +K_\pm^\bomega(\x+\bomega t,\,\p)\Big\}\,\psi\,dt \ . \ee
The relation
 $a(\bomega,\,\x)=\lambda_+^\bomega(\x)-\lambda_-^\bomega(\x)$ gives
(\ref{a1}) as a weak limit. Moreover, we have
 $\e{-\i\u\x}\,\Omega_\pm^*\,\e{\i\u\x}\to\e{\i\lambda_\pm^\bomega(\x)}$
strongly, and the strong limit in (\ref{a1}) is obtained from the
decomposition
 \be S-\e{\i a}= \Omega_+^*\,
 \Big(\Omega_- -\e{-\i\lambda_-^\bomega}\Big) + \Omega_+^*\,
 \Big(\e{-\i\lambda_+^\bomega}-\Omega_+\Big)\,\e{\i a} \ . \ee
For the limit of $\Omega_+\,$, we are applying (\ref{cooka}) to
 $\e{\i a}\,\psi\in W^2$ instead of $\psi$, which requires two additional
arguments: First, the regularization of $A_0$ is not done with
$\psi=g(\p)\psi$, but with
 $\e{\i a}\,\psi=(H_0+\i)^{-1}\Big((H_0+\i)\,\e{\i a}\,\psi\Big)$. Second,
in the proof of the nonpropagation property, $(\e{\i a}\,\psi)^\wedge$ does
not have the desired compact support. But $a(\bomega,\,\x)$ is constant in
the direction $\bomega$, thus the support of $\hat\psi$ is enlarged only
orthogonal to $\bomega$, and remains bounded in the direction of $\bomega$.
(Use only the directional derivative $\bomega\cdot\nabla_\p$ in the proof
of the nonstationary phase estimate \cite[Thm.~1.8]{th}.)

Now suppose that $S$ is known (it is invariant under short-range gauge
transformations),
thus the absolute phase of its high-energy limit is known. Then $\B$ is
reconstructed by Thm.~\ref{The}, and any corresponding short-range $\A$ gives
$a(\bomega,\,\x)$. Since $K_\pm^\bomega(\x,\,\p)$ can be computed
from $\B$, the X-ray transform of $A_0$ is obtained from (\ref{a1}). $A_0$ is
reconstructed in the second step according to \cite{he}, at least under
stronger regularity assumptions. In
the general case, the potential is regularized by translated test functions
\cite{ew2}, or it is considered in $\S'$ \cite{wjde}. \mybox

The main difference to Arians' original proof \cite{ar1} is the adaptive
gauge transformation $\A_\pm^\bomega=\A+\grad\lambda_\pm^\bomega\,$: since
$\bomega\cdot\A_\pm^\bomega(\x)\equiv0$, the
limit (\ref{a5}) is read off easily from (\ref{cooka}) after showing the
uniform bound, and the expression (\ref{a6}) for $K_\pm^\bomega(\x,\,\p)$ is
obtained from $H_\pm^\bomega-H_0$ without calculation. In \cite{ard},
magnetic fields with compact
support and nonvanishing flux are considered as well, by employing a family
of transversal gauges with adapted reference point. Cf.~Cor.~\ref{Cc}.
Analogously we have

\begin{rmk}[Generalization (Arians)] \label{Ra}
Suppose $\nu=2$ and $A_0\,$, $\B$ satisfy the assumptions of Thm.~\ref{Ta},
except the flux is not vanishing. Then (\ref{a1}) and the proof remain
valid, if $\A(\x)$ decays integrably in the half-plane $\bomega\cdot\x>0$
and in a sector around $-\bomega$. When $\A$ is fixed, we may consider a
family of gauge transformations according to Cor.~\ref{Cc} to satisfy the
decay requirements, and the right hand side of (\ref{a1}) is modified.
\end{rmk}

As noted in \cite{ar1, ard}, the uniform estimate of the Cook integral
(\ref{cooka}) and Remark~\ref{Ra} give error bounds for the high-energy
limit of $S$ according to Thm.~\ref{The}:

\begin{cor}[Error Bounds (Arians)] \label{Ce}
$1.$ Under the short-range assumptions of Thm.~\eref{Ta}, the limit
\emph{(\ref{hes})} has an explicit error bound for
$\hat\psi\in C_0^\infty\,$, which is of the form
 \be \e{-\i\u\x}\,S\,\e{\i\u\x}\,\psi
 =\e{\i a(\bomega,\,\x)}\,\psi+\O(1/u) \ . \ee

$2.$ If $\nu=2$ and the flux of $\B$ is not vanishing, then an analogous
weak estimate remains valid for $\hat\phi,\,\hat\psi\in C_0^\infty$ and
medium-range $\A$ with $\A(\x)=\O(1/|\x|)$:
 \be \Big(\phi,\,\e{-\i\u\x}\,S\,\e{\i\u\x}\,\psi\Big)
 =\Big(\phi,\,\e{\i a(\bomega,\,\x)}\,\psi\Big)+\O(1/u) \ . \ee
\end{cor}

Error bounds for (\ref{a1}) would require stronger decay assumptions on
$A_0$ \cite{ew2}, or stronger regularity assumptions \cite{nfo}. The
right hand side of (\ref{a1}) contains a multiplication operator times
$\p$. In the short-range case, it can be rewritten according to
 \be \int_{-\infty}^0 \A_-^\bomega(\x+\bomega t)\,dt 
 + \int_0^\infty \A_+^\bomega(\x+\bomega t)\,dt
 = \int_{-\infty}^\infty t\bomega\times\B(\x+\bomega t)\,dt
 = \nabla_\bomega\, a(\bomega,\,\x) \ . \ee
Only the last expression remains meaningful, if $\B$ is smooth but does
not not decay faster than $|\x|^{-2}$: an asymptotic expansion of
 $\e{-\i\u\x}\,S\,\e{\i\u\x}$ in powers of $1/u$ has been obtained
in \cite{nfo} for $\B\in C^\infty$, 
$\partial^\alpha \B(\x)=\O(|\x|^{-\mu-|\alpha|})$, $\mu>3/2$.

In the special case of $\B=0$ and $\A=0$, thus $H=H_0+A_0\,$, the uniform
estimate of (\ref{cooka}) together with
$S-1=\Omega_+^*(\Omega_--\Omega_+)$ and the strong limit of $\Omega_+^*$
yield
 \be\label{ew} \lim_{u\to\infty} \,\dfrac{u}{m}\,\Big(\e{-\i\u\x} \,
 S\, \e{\i\u\x}-1\Big) \,\psi \,=\,
 -\i \int_{-\infty}^\infty A_0 (\x+\bomega t)\,\psi \,dt \ee
for $\hat\psi\in C_0^\infty$. This limit was obtained by Enss and Weder in a
series of papers, cf.~\cite{ew1, ew2},
introducing the time-dependent geometric method, and including also the cases
of $N$-particle scattering, inverse two cluster scattering, and long-range
electrostatic potentials. Error bounds for the weak formulation of
(\ref{ew}) were established in \cite{ew2} under stronger decay assumptions.
The strong limit is due to \cite{wjdip}.

\section{Concluding Remarks} \label{Sphint}
A geometric interpretation of the scattering process at high energies is
given in \cite{ew1, eji}, cf.~Sec.~\ref{subhe}. Here we shall discuss the
implications of gauge invariance and the question of measurable
quantities in a scattering process, as well as the possible application
of the inverse scattering problem.

\subsection{Gauge Invariance}  \label{subgi}
The vector potential $\A$ is determined only up to a gradient by the
magnetic field $\B$ (and by additional discrete values of fluxes, when the
domain is multiply connected). It is not eliminated easily from the
Schr\"odinger equation $\i\dot\psi=H\psi$. (For the nonlinear hydrodynamic
formalism, cf.~\cite{swcsr, rab, ptab}.) The Schr\"odinger equation or Pauli
equation is invariant under the simultaneous gauge transformation of
$\A$ and $\psi$,
 \be \label{gtap} \A\to\A'=\A+\grad\lambda \qquad
 \psi\to\psi'=\e{\i\lambda}\,\psi \ . \ee
(If the electromagnetic field was time-dependent, we would have
$A_0'=A_0-\dot\lambda$ in addition.)
One interpretation is, that the transformation of $\A$ comes from
$\curl\A=\B$, and it needs to be compensated for by the transformation of
$\psi$. Following Weyl, this argument can be reversed: if we assume that
the local phase of the wave function $\psi$ is not observable, the theory
should be invariant under a local $U(1)$ transformation, and $\psi$ must be
coupled to a vector potential. It is assumed in general that $\A$ and $\A'$
describe the same magnetic field, and that all observable physical effects
are independent of the chosen gauge. Thus an electron in an electromagnetic
field is described by an equivalence class of pairs $(\A,\,\psi)$, where
$(\A,\,\psi)\sim(\A',\,\psi')$ iff there is a $\lambda$ with (\ref{gtap}).

Gauge invariance implies that not every self-adjoint operator corresponds
to a physical observable: Suppose that the self-adjoint operator
$F=F(\p,\,\A(\x),\,\dots)$ is constructed from a function
$f(\p,\,\A(\x),\,\dots)$ by some quantization procedure (since the ordinary
functional calculus does not apply due to $[\x,\,\p]\neq0$). Then the
expectation value $(\psi,\,F\psi)$ is gauge-invariant, i.e.,
it depends only on the equivalence class of $(\A,\,\psi)$, iff
 \be \e{\i\lambda(\x)}\,F(\p,\,\A(\x),\,\dots)\,\e{-\i\lambda(\x)}
 = F(\p,\,\A(\x)+\grad\lambda(\x),\,\dots) \ . \ee
This restriction is a ``superselection rule'' in the general sense of
\cite{swcsr}.
At least if $f$ is polynomial in $\p$, this means that it depends on
$\p$ and $\A$ only in the combination $\p-\A$. Examples of nonobservable
operators are $\A(\x)$, the canonical momentum operator $\p$, the free
Hamiltonian $H_0=\frac1{2m}\p^2$, and the canonical angular momentum
$\mathbf{L}=\x\times\p$. The following operators are among the observables:
$\x$, $A_0(\x)$, $\B(\x)$, the kinetic momentum $m\dot\x=\p-\A$, the kinetic
energy $\frac1{2m}(\p-\A)^2$, the Hamiltonian $H$, and the kinetic angular
momentum $\x\times m\dot\x=\x\times(\p-\A)=\mathbf{L}-\x\times\A$. See
\cite{swcsr, wjm2} for a discussion of vector potentials and gauge
invariance in the context of the Aharanov--Bohm effect \cite{ptab}.

\subsection{The Scattering Cross Section}  \label{subscs}
When $\B=0$, or for the time evolution of asymptotic configurations, it is
natural to set $\A=0$ by convention. In the
scattering theory with medium-range vector potentials, we must assume that
$\A$ and $\A'$ describe the same physical system, as soon as
$\curl\A=\curl\A'$. Then $\A'-\A$ need not be short-range, and the scattering
operator $S$ is not gauge-invariant, but it transforms according to
(\ref{gts}). Given a scattering state $\psi\in\Ran(\Omega_+)$, the particle
is found in a cone $\mathcal{C}$ for $t\to+\infty$ with probability
 \be \label{cone} \lim_{t\to+\infty}
 \Big\|\,F(\x\in\mathcal{C})\,\e{-\i Ht}\,\psi\,\Big\|^{\,2}
 \,=\, \Big\|\,F(\p\in\mathcal{C})\,\Omega_+^*\,\psi\,\Big\|^{\,2} \ee
according to Dollard \cite[Thm.~IX.31]{rs2}. Now
$\Omega_+^{\prime*}\psi'=\e{\i\Lambda(\p)}\,\Omega_+^*\psi$ by (\ref{gts}),
thus this number is gauge-invariant. By the correspondence between subsets of
$S^{\nu-1}$ and cones in $\R^\nu$ (with apex $0$), (\ref{cone}) defines a
measure on $S^{\nu-1}$. The differential cross section
$d\sigma/d\omega$ for incident momentum $\q$ is obtained when
$\phi=\Omega_-^*\psi$ is approaching a plane wave, rescaled such that its
Fourier transform $\hat\phi(\p)$ is approaching
 ``$\sqrt{\delta(p_1-q)}\,\delta(p_2)\,\delta(p_3)$'' when
$\q=(q,\,0,\,0)^\tr$. If the momentum support of an incoming asymptotic
configuration $\phi$ is concentrated at $\p\approx\q$, or at
$\p\in\q[0,\,\infty)$, we have
 \be S' \phi = \e{\i\Lambda(\p)}\,S\,\e{-\i\Lambda(-\p)}\,\phi
 \approx \e{\i\Lambda(\p)}\,S\,\e{-\i\Lambda(-\q)}\,\phi
 = \e{\i\Lambda(\p)-\i\Lambda(-\q)}\,S\,\phi \ . \ee
The phase factor in momentum space does not influence the probability of
finding the outgoing state in a cone, thus $d\sigma/d\omega$ is
gauge-invariant. By the same argument, i.e., replacing $\Lambda$ with $a$,
we may compute it from the gauge-invariant modified scattering operator
$S^J$, which was defined in (\ref{sjs}). See also \cite{tss, ryab}.

\subsection{The Phase of the Scattering  Amplitude}  \label{subpa}
In short-range scattering theory, the differential cross section is
$\dfrac{d\sigma}{d\omega}=\Big|f_\q(\bomega)\Big|^2$, where $f_\q(\bomega)$
is the scattering amplitude for incident momentum $\q$ and outgoing direction
$\bomega$. It is obtained from the $T$-matrix, i.e., the integral kernel of
$S-1$ on the energy shell. Probably this relation remains valid in the
medium-range situation,
although the latter kernel will be more singular on the diagonal \cite{yhs}.
In scattering experiments, usually the differential cross section is observed
for an incident beam of particles, which is modeled as a plane wave.
Information on the phase of the scattering amplitude is not available
directly, but it is required for solving the inverse scattering problem,
e.g., with (\ref{ew}).

For a small, central, scalar potential in $\R^3$, $f_\q$ can be reconstructed
from $\Big|f_\q(\bomega)\Big|^2$ by employing the unitarity of $S$, see
\cite[Sec.~V.6.D]{rs1} and the references in \cite{ksds}. If this approach
is extended to $\R^2$, it will work for rotationally symmetric $\B$ as well.
Phase information
would be available experimentally, if it was possible to localize the
incoming particles more precisely. It could be reconstructed as well, if the
location of the unknown scatterer is kept fixed, and the location of a known
additional potential is varied while measuring the cross sections
\cite{ksds}. In some cases, phase information is obtained from interference
between the scattered beam and a coherent reference beam \cite{ptab}.

\subsection{Two-Particle Scattering}  \label{sub2}
If two nonrelativistic particles are interacting via a pair potential
$A_0(\x_2-\x_1)$, their relative motion is equivalent to an external field
problem for one particle with the reduced mass $m=\frac{m_1m_2}{m_1+m_2}\,$.
The pair potential has a physical justification only if it is
a central potential, or if, say, $m_2\ll m_1\,$: Suppose particle~1 is a
molecule with a dipole field given by $A_0(\x_2-\x_1)$, and particle~2 is an
electron. The molecule will be rotated by interacting with the
electron, but the corresponding rotation of $A_0$ is neglected in the model.
This simplification is justified if $m_1\gg m_2\,$, and in this case we might
assume as well that the molecule is generating a magnetic field.
If the orientation of particle~1 is unknown, the scattering
cross section may be defined by averaging over the orientations. But the
phase information required for solving the inverse problem is unlikely to
be recovered.

\subsection{Inverse Scattering and Error Bounds}  \label{subie}
The high-energy asymptotics (\ref{hes}), (\ref{a1}), (\ref{ew}) might be
applied independently from inverse scattering. But consider the inverse
scattering problem for a particle in an external electromagnetic field,
or for two-particle scattering under the restrictions from Sec.~\ref{sub2}.
The uniqueness of the solution is interesting from a theoretical point of
view, because in the analogous situation of particle physics, the models
are based mainly on scattering data. In our situation, two additional
questions must be addressed, before a field can be reconstructed
from a scattering experiment: the problem of obtaining phase information,
cf.~Sec.~\ref{subpa}, and the effects of a high but finite energy. When
a-priori bounds on $A_0\,$, $\B$, and $\A$ are given, it is possible to
estimate the approximation error in (\ref{hes}) according to Cor.~\ref{Ce}.
Thus we can check in principle, if the required energy is available in the
experimental setup, and if the nonrelativistic model makes sense for this
high energy (note also that the scattering operators for the Pauli- and
Dirac equations (positive energy) coincide if $A_0=0$,
cf.~\cite{th, wjm2}).

If the high-energy limit was observed for a suitable
family of states $\psi$, we could obtain the required X-ray transforms as
multiplication operators. At a high but finite energy, we do not get the
operator of multiplication with an approximate X-ray transform, and it is
not clear how to obtain the latter. (If the X-ray transform was obtained
approximately, we could apply regularization techniques to provide an
approximate inversion of the X-ray transform, whose exact inversion is
ill-posed.) For $A_0$ of compact support, it is suggested in \cite{ttrr}
to consider (\ref{ew}) for a single chosen $\psi$. Or we might
specify an inversion procedure for the X-ray transform and apply it to
the high-energy asymptotics, to check if the resulting composition of
operators is converging.
It may be possible as well, to obtain $A_0$ at a lower energy by a recursive
approach, or by considering more terms of the asymptotic expansion.

{
Papers by Arians, Enss, or Jung are available from
\href{http://www.iram.rwth-aachen.de}{http://www.iram.rwth-aachen.de}
(except for \cite{ard}).}

\section*{Corresponding Results on Relativistic Scattering\rule{0mm}{20mm}}
This appendix of the preprint will not be part of the published paper.
It should be considered as a summarizing preview of \cite{wjm2}.
\renewcommand{\theequation}{A\arabic{equation}} \setcounter{equation}{0}
\renewcommand{\thethm}{A\arabic{thm}} \setcounter{thm}{0}
\bmdefine{\balpha}{\alpha} \def\sign{\mathop{\rm sign}\nolimits}
\newcommand{\twovec}[2]%
 {\left(\begin{array}{c}{#1}\\[0mm]{#2}\end{array}\right) }

The Dirac operators are matrix-valued operators of the form
 $H_0=\balpha\cdot\p+\beta m$ and $H=\balpha\cdot(\p-\A)+\beta m+A_0\,$.
The vector potential $\A$ shall be of medium range. The scalar potential
shall be continuous except for a finite number of local singularities, where
$|A_0(\x)|\le c|\x-\x_j|^{-\mu}$ with $\mu<1$, and decay integrably.
The scattering operator $S$ is decomposed according to the subspaces of
positive or negative energy, and we will need the Newton-Wigner position
operator $\tilde\x$ \cite{th}:

\begin{thm}[High-Energy Asymptotics and Inverse Scattering] 
Suppose that $\B$ is a magnetic field of medium decay in $\R^2$ or $\R^3$,
$\A$ is any medium-range vector potential with $\curl\A=\B$ in $\S'$, and
$\A_0$ is a short-range electrostatic potential.

\emph{1.}  The wave operators $\Omega_\pm$ and the scattering operator $S$
for the Dirac equation exist.  Consider also a gauge transformation
$\A'=\A+\grad\lambda$ and $\Lambda(\x)=\lim_{r\to\infty} \lambda(r\x)$,
and denote the operators corresponding
to $\A'$ by $H'$, $\Omega_\pm'$, $S'$. They obey the gauge transformation
formula
\be\label{gtd}  \Omega_\pm' \,=\,
 \e{\i\lambda(\x)} \,\Omega_\pm\, \e{-\i\Lambda(\pm\p\,\sign(H_0))} \qquad
 S_\pm' \,=\, \e{\i\Lambda(\pm\p)} \,S_\pm\, \e{-\i\Lambda(\mp\p)} \ , \ee
where $\S_\pm$ denotes the restriction of $S$ to the subspace of
positive/negative energy.

\emph{2.} Consider translations in momentum space by $\u=u\bomega$,
$\bomega\in S^{\nu-1}$. Denote by $S_\pm$ the restriction of $S$ to the
subspace $P_\pm\H$ of positive/negative energy. With the
Newton-Wigner position operator $\tilde\x$ we have the high-energy limit
\be\label{hed}\
 \slim_{u\to\infty} \,\e{-\i\u\tilde\x} \,S_\pm\, \e{\i\u\tilde\x} \,=\,
 \exp\Big\{ -\i\int_{-\infty}^\infty \Big( A_0 \mp \bomega\cdot\A \Big)
 (\tilde\x\pm\bomega t) \,dt \Big\} \ . \ee
Both $A_0$ and $\B$ are reconstructed from the relative phase of the
high-energy limit of $S_+$. Analogous results hold for
the Klein--Gordon equation.
\end{thm}

\renewcommand{\thepage}{A2}%
\textbf{Sketch of the proof}: 1. The Cook integral is estimated by
employing the special gauge $\A=\A^r+\A^s$ and applying the technique of
\cite{ltr} to $\A^r$. We have
 $\A^r\cdot\balpha=\A^r\cdot\p/H_0+\A^r\cdot(\balpha-\p/H_0)$. The first
term is written as $\A^r\cdot\p=\G\cdot\mathbf{L}$, and the second term
is controlled with partial integration, since it is oscillating due to
$\{\balpha-\p/H_0\,,\,H_0\}=0$. The usual density argument gives the
existence of $\Omega_\pm\,$. The gauge transformation formula was proved
in \cite{wjdip}.

2. The high-energy limit of \cite{wjde} is extended in two directions:
Medium-range vector potentials are included by estimating the time
evolution analogously to the direct problem. Local singularities of
$A_0$ are treated with a density argument and variable cutoff functions.

The corresponding results for the Klein--Gordon equation are obtained
analogously by writing it as a first-order system in the Foldy--Wouthuysen
representation and by employing the Dyson expansion. Now $A_0$ shall be
bounded and $\|A_0(x){\sqrt{\p^2+m^2}}^{\,-1}\|<1$. The proof was given in
\cite{eji} under more restrictive conditions, it is extended again with
$\A^r\cdot\p=\G\cdot\mathbf{L}$. \mybox

Consider a compact set $K\subset\R^\nu$ and a Hamiltonian $H_K$ in
$L^2(\R^\nu\setminus K)$. Obstacle scattering means to deal with the wave
operators
\be \label{omo} \Omega_\pm \,:=\,
 \slim_{t\to\pm\infty}\,\e{\i H_Kt}J\,\e{-iH_0t} \ , \ee
where $J$ is the natural projection from $L^2(\R^\nu)$ onto
$L^2(\R^\nu\setminus K)$. Nonrelativistic inverse obstacle scattering using
high-energy limits was discussed by Nicoleau \cite{nab} and
Weder \cite{wab}. We shall consider the Dirac equation in two cases:

i) The compact set $K\subset\R^\nu$ is convex and space-reflection
symmetric, i.e., $K=-K$. $\balpha\cdot\p$ is a symmetric operator on
$C_0^\infty(\R^\nu\setminus K)$, and the deficiency indices of its closure
are equal, since it is anticommuting with the unitary involution
$(R\psi)(\x)=\psi(-\x)$. Fix any self-adjoint extension and employ the
Kato-Rellich Theorem to define
 $H_K:=\balpha\cdot\p+(\beta m-\balpha\cdot\A+A_0)$ when $A_0$ is bounded.

ii) In the case of $K=\{\0\}\subset\R^2$, stronger singularities of $\A$
at $\x=\0$ are permitted: the
magnetic field shall be of the form $\B(\x)=\B_m(\x)+\Phi_*\delta(\x)$ with
$\B_m$ of medium decay and $\Phi_*\in\R$. The vector potential has a
decomposition $\A(\x)=\A_m(\x)+\A_*(\x)$, where $\A_m$ is of medium range
and $\curl\A_m=\B_m$.
\be \A_*(\x) \,:=\, \frac{\Phi_*}{2\pi} \,|\x|^{-2}\, \twovec{-x_2}{x_1} \ee
satisfies $\curl\A_*=\Phi_*\delta$, and it behaves as a vector potential of
medium range for $|\x|\to\infty$. Now $\A_*(\x)$ is odd, thus
 $\balpha\cdot(\p-\A_*)$ is a
symmetric operator with equal deficiency indices. Fix any self-adjoint
extension and include $\beta m-\balpha\cdot\A_m+A_0$ with Kato-Rellich.

\begin{thm}[Obstacle Scattering] \label{To}
Consider an obstacle $K\subset\R^\nu$ and a Dirac operator $H_K$
satisfying Assumption~\emph{i)} or~\emph{ii)} above.

\emph{1.} The wave operators \emph{(\ref{omo})} exist and are isometric,
they transform under a change of gauge according to \emph{(\ref{gtd})}.

\emph{2.} For $\bomega\in S^{\nu-1}$ and all $\psi\in L^2(\R^\nu,\,\C^\mu)$
with $\tilde\x$-support outside of the cylinder $K+\R\,\bomega$,
i.e., $F(\tilde\x\in K+\R\,\bomega)\psi=0$, we have the high-energy limit
 \be\label{o}\wlim_{u\to\infty}\,\e{-\i\u\tilde\x}\,S_+\,\e{\i\u\tilde\x}
 \,\psi\,=\,
 \exp\Big\{ -\i\int_{-\infty}^\infty \Big( A_0 - \bomega\cdot\A \Big)
 (\tilde\x+\bomega t) \,dt \Big\} \,\psi \ . \ee

\emph{3.} For $\x\in\R^\nu\setminus K$, the electrostatic potential
$A_0(\x)$ and the magnetic field $\B(\x)$ are reconstructed from the
relative phase in \emph{(\ref{o})}. If
$\nu=2$ and $K\neq\{\0\}$, we need the additional assumption that both
decay faster than any power.

\emph{4.} If $\nu=2$ and $\B$ has a finite flux $\Phi$, then $\Phi$ is
reconstructed modulo $2\pi$ from the relative phase in \emph{(\ref{o})}.
\end{thm}

If the absolute phase was observable, then $\Phi$ could be reconstructed
modulo $4\pi$ \cite{nab, wab}.
Note that the obstacle is assumed to be known, and only the fields are
reconstructed. Different self-adjoint choices of $H_K$ are not distinguished
in the high-energy limit.  The stronger decay assumptions in item~3 are
required by the Support Theorem for the X-ray transform \cite{he}. In the
case ii), items~3 and~4 mean that $\B_m$ is reconstructed uniquely, and
$\Phi_*$ is reconstructed modulo $2\pi$. If $\nu=2$ and the flux of $\B$
on $K$ is nonzero, it is influencing the particle outside of $K$ via
the vector potential (Aharanov--Bohm effect).

\renewcommand{\thepage}{A3}%
\textbf{Sketch of the proof}: The Cook integral is estimated analogously
to the case without obstacle, by replacing the projector $J$ with a
smooth cutoff function $\chi(\x)$. When interchanging the limits $u\to\infty$
and $t\to\pm\infty$, the free time evolution is estimated by introducing
a momentum cutoff $f(\p/u)$ as in \cite{wab}, since the states
do not have compact momentum support. For the high-energy limit at a
finite time, the Dyson expansion does not apply when $K\neq\{\0\}$, since
$J$ is not injective. Introducing the Dirac operator $H$ in
$L^2(\R^\nu)$, consider the decomposition
 \ban && \e{\i H_Kt}\,\chi\,\e{-\i H_0t}\,\psi
 \;=\; \Big(\e{\i H_Kt}\,\chi\,\e{-\i Ht}\Big)\,
 \Big(\e{\i Ht}\,\e{-\i H_0t}\Big)\,\psi \\[1mm]
 &=& \chi\,\e{\i Ht}\,\e{-\i H_0t}\,\psi
 \;+\; \i\int_0^t\e{\i H_Ks}\,(H_K\chi-\chi H)\,\e{\i H(t-s)}\,\e{-\i H_0t}
 \,\psi\,ds \ .\quad\rule{0mm}{0mm} \label{ax} \ean
The high-energy asymptotics of the time evolution are known for $H_0$ and
$H$ but not for $H_K$. After performing the known limits, the integral is
seen to vanish because of the support properties of $\psi$ and $\chi$,
and the limit of the first term in (\ref{ax}) remains. \mybox

\end{document}